\documentclass[prb,aps,nobibnotes,twocolumn,superscriptaddress,10pt,longbibliography]{revtex4-1}
\usepackage[utf8]{inputenc}
\usepackage[T1]{fontenc}
\usepackage[
  top=8mm,
  bottom=7mm,
  left=1.5cm,
  right=1.5cm,
  headheight=14pt,
  includehead,includefoot,
  heightrounded, 
]{geometry} 
\usepackage{color}
\usepackage{amsmath}
\usepackage{amsfonts}
\usepackage{hyperref}
\usepackage{comment}
\usepackage{multirow}
\usepackage{amsthm}
\usepackage{bm}
\usepackage{float}
\usepackage{fancyhdr}
\usepackage{filecontents}
\hypersetup{
  colorlinks = true,
  allcolors= blue,
}
\usepackage[percent]{overpic}
\usepackage{tikz,pgfplots}
\usepackage{blkarray}
\usepackage[normalem]{ulem}
\usepackage{bibunits}
\usepackage{etoolbox}

\makeatother
\DeclareUnicodeCharacter{300}{à}

\DeclareMathAlphabet{\mathsfit}{\encodingdefault}{\sfdefault}{m}{sl}
\SetMathAlphabet{\mathsfit}{bold}{\encodingdefault}{\sfdefault}{bx}{sl}
\newcommand{\tens}[1]{\bm{\mathsfit{#1}}}
\newcommand{\tenscomp}[1]{\mathsfit{#1}}
\makeatletter
\fancypagestyle{footmark}{
  \fancyfoot[L]{\footmark}
}
\newcommand\markfoot[1]{\gdef\footmark{#1}\thispagestyle{footmark}}
\makeatother

\makeatletter
\renewcommand*{\@fnsymbol}[1]{\ifcase#1\else\@arabic{\numexpr#1\relax}\fi}
\makeatother
\makeatletter
\newcommand*{\newbibstartnumber}[1]{%
  \apptocmd{\thebibliography}{%
    \global\c@NAT@ctr #1\relax
    \addtocounter{NAT@ctr}{-1}%
  }{}{}%
}
\makeatother
\begin{document}
\title{\sf \textbf{\Huge Unified theory of thermal transport in\hspace*{24mm}~\\[2mm] crystals and disordered solids\hspace*{60mm}~}}
\author{\sf {{Michele Simoncelli$^{1}$}}}
\author{\sf {{Nicola Marzari$^{1}$}}}
\author{\sf {{Francesco Mauri$^{2\star}$\hspace{95mm}~}}}
\maketitle
\thispagestyle{empty}

{\sf\textbf{
\hspace*{-5mm}Crystals and glasses exhibit fundamentally different heat conduction mechanisms: the periodicity of crystals allows for the excitation of propagating vibrational waves that carry heat, as first discussed by Peierls\cite{peierls1929kinetischen};
 in glasses, the lack of periodicity breaks Peierls' picture and heat is mainly carried by the coupling of vibrational modes, often
described by a harmonic theory introduced by Allen and Feldman\cite{allen1989thermal}.
Anharmonicity or disorder are thus the limiting factors for thermal conductivity in crystals or glasses; hitherto, no transport equation has been able to account for both.
Here, we derive such equation, resulting in a thermal conductivity that reduces to the Peierls and Allen-Feldman limits, respectively, in anharmonic-and-ordered or harmonic-and-disordered solids, while also covering the  
intermediate regimes where both effects are relevant. 
This approach also solves the long-standing problem of 
accurately predicting the thermal properties of crystals with ultralow or glass-like
thermal conductivity\cite{sun2010lattice,li2015ultralow,wang2018cation,lee2017ultralow,chen2015twisting,voneshen2013suppression,lory2017direct,PhysRevB.96.214202}, as we show with an application to a thermoelectric material representative of this class.
}}

\markfoot{\vspace*{-7mm}\hrule\vspace*{1mm}\sf\footnotesize{ $^{1}$Theory and Simulation of Materials (THEOS) and National Centre for Computational Design and Discovery of Novel Materials (MARVEL), {\'E}cole Polytechnique F{\'e}d{\'e}rale de Lausanne, Lausanne, Switzerland. $^{2}$Dipartimento di Fisica, Universit{\`a} di Roma La Sapienza, Piazzale Aldo Moro 5, I-00185 Roma, Italy. $^\star$e-mail: \textcolor{blue}{francesco.mauri@uniroma1.it}}}

In 1929 Peierls\cite{peierls1929kinetischen} developed a semi-classical theory for heat conduction in terms of a Boltzmann transport equation (BTE) for  propagating phonon wave packets.
Nowadays, modern algorithms and computing systems allow to solve its linearized form (LBTE) either approximately (in the so-called single mode approximation (SMA)\cite{PhysRevLett.106.045901}) or exactly, using iterative\cite{omini1995iterative,carrete2017almabte}, variational\cite{fugallo2013ab}, or exact diagonalization\cite{PhysRevLett.110.265506,cepellotti2016thermal} methods;  its accuracy has been highlighted in many studies\cite{fugallo2013ab,cepellotti2016thermal,cepellotti2015phonon,PhysRevLett.110.265506}. 
Nevertheless, these cases are characterized by having few, well-separated phonon branches and anharmonicity-limited thermal conductivity; we will refer to these in the following as ``simple'' crystals. 
In 1963 Hardy was able to express the thermal conductivity
in terms of the phonon velocity operator\cite{hardy1963energy} and showed that 
its diagonal elements are the phonon group velocities entering the Peierls' BTE, while the  off-diagonal terms, missing from it, are actually negligible in simple crystals\cite{hardy1963energy}.
In 1989 Allen and Feldman\cite{allen1989thermal} envisioned that these off-diagonal elements, neglected so far, 
could become dominant in disordered regimes,
where Peierls'  picture breaks down due to the impossibility of defining phonons and group velocities. 
As a consequence, a harmonic theory of thermal transport in glasses was introduced, where disorder limits thermal conductivity and heat is carried by couplings of vibrational modes arising from the off-diagonal elements of the velocity operator (diffusons and locons\cite{allen1999diffusons,lv2016non}).
Recently, it has been argued that the diffuson conduction mechanism, typical of glasses, can also emerge in a third class of materials, termed ``complex'' crystals\cite{chen2015twisting}, characterized by large unit cells and many quasi-degenerate phonon branches, where it coexists with phonon transport.
Conversely, crystal-like propagation mechanisms have been suggested also for glasses (propagons\cite{allen1999diffusons}) --- albeit without a formal justification --- in order to explain the experimental results\cite{allen1999diffusons}.

Here, we  show that a more general theory exists, encompassing the emergence and coexistence of all these excitations, and reducing to the BTE in the semi-classical limit of a simple crystal, or to the Allen-Feldman formulation in the case of a harmonic glass.

We start by considering a crystal 
in the Born-Oppenheimer approximation, where we add perturbatively to a harmonic Hamiltonian\cite{ziman1960electrons} $\mathrm{H}^{\rm har}$ a collision term $\mathrm{H}^{\rm col}$ accounting for anharmonicity\cite{ziman1960electrons} and isotopic disorder\cite{tamura1983isotope}.
Heat transfer is driven by a temperature gradient in real space (\textit{i.e.} a space-dependent lattice vibrational energy) and 
in order to derive a more general transport equation we want to track excitations both in real and Fourier space.
With this goal (see Methods for details) we introduce the bosonic operator $ \mathrm{a}(\bm{q})_{b\alpha}$ that annihilates  vibrations with wave-vector $\bm{q}$ for all atoms $b$ in direction $\alpha$ 
(the Einstein summation convention is implied over repeated indexes)
\begin{equation}\label{eq:def_op_a}
  \mathrm{a}(\bm{q})_{b\alpha}{=}\tfrac{1}{\sqrt{2 \hbar}}\hspace*{-0.8mm} \Big(\hspace*{-1.5mm}\sqrt[4]{\hspace*{-0.7mm}\tenscomp{D}^{\hspace*{-0.1mm}{-}1}\hspace*{-0.6mm}(\bm{q})}_{\hspace*{-0.5mm}b\alpha,b'\hspace*{-0.5mm}\alpha'}\hspace*{-0.5mm}{\mathrm{P}}(\bm{q})_{\hspace*{-0.4mm}b'\hspace*{-0.5mm}\alpha'} {-}i \hspace*{-0.5mm}\sqrt[4]{\hspace*{-0.7mm}\tenscomp{D}(\bm{q})}_{\hspace*{-0.5mm}b\alpha\hspace*{-0.2mm},\hspace*{-0.2mm}b'\hspace*{-0.5mm}\alpha'}\hspace*{-0.5mm}{\mathrm{U}}^\dagger\hspace*{-0.5mm}(\bm{q})_{\hspace*{-0.4mm}b'\hspace*{-0.5mm}\alpha'}\hspace*{-1.3mm}\Big)\hspace*{-0.5mm}.
\end{equation}
Here ${\mathrm{U} }(\bm{q})_{b\alpha}{=}\sqrt{{{M_b}}/{{N_{\rm c}} }}\sum_{\bm{R}} \mathrm{u}(\bm{R})_{b\alpha}e^{{+}i\bm{q}\cdot \bm{R}}$ and ${\mathrm{P}}(\bm{q})_{b\alpha}{=}({\sqrt{M_b N_{\rm c}} })^{-1}\sum_{\bm{R}} \mathrm{p}(\bm{R})_{b\alpha}e^{-i\bm{q}\cdot\bm{R}}$ are the Fourier representations of the displacement and momentum operators $\mathrm{u}(\bm{R})_{b\alpha}$ and $\mathrm{p}(\bm{R})_{b\alpha}$ of atom $b$ in unit cell $\bm{R}$, with mass $M_b$ and in a lattice containing $N_{\rm c}$ unit cells, and 
$\tens{D}(\bm{q})$ is the dynamical matrix of the crystal with eigenvalues $\omega^2(\bm{q})_s$ ($s$ labels the $3{\cdot}N_{\rm at}$ phonon branches, $N_{\rm at}$ being the number of atoms in the unit cell). 
The usual commutation relations 
$[\mathrm{a}(\bm{q})_{b\alpha},\mathrm{a}^\dagger(\bm{q}')_{b'\hspace*{-0.5mm}\alpha'}]{=}\delta_{\bm{q,q'}}\delta_{b,b'}\delta_{\alpha,\alpha'}$ 
are satisfied, and 
in terms of these operators  
$\mathrm{H}^{\rm har}$ assumes the simple form:
\begin{equation}
\begin{split}
    &\mathrm{H}^{\rm har}{=} \hbar \hspace{-0.5mm}\sum\limits_{\bm{q}} 
\hspace{-1mm}   \sqrt{\hspace{-0.5mm}\tenscomp{D}(\bm{q})}_{b\alpha,b'\hspace{-0.5mm}\alpha'}
       \big(\mathrm{a}^{\dagger}(\bm{q})_{b\alpha}\mathrm{a}(\bm{q})_{b'\hspace{-0.5mm}\alpha'}{+}\frac{1}{2}\delta_{b,b'}\delta_{\alpha,\alpha'\hspace{-0.5mm}}\big).\hspace*{7mm}
  \label{eq:full_H}
\end{split}
\raisetag{8mm}
\end{equation}
When a temperature gradient is imposed with couplings to external baths,
the system's state --- described by the density matrix $\rho$ --- undergoes an irreversible Markovian evolution described by the 
master equation\cite{gebauer2004kinetic} 
\begin{equation}
	\frac{{\partial} {\rho}}{\partial t}(t)+\frac{i}{\hbar}\Big[\mathrm{H}^{\rm har},{\rho}(t)\Big]=\frac{\partial \rho(t) }{\partial  t }\bigg|_{\mathrm{H}^{\rm col}};
\label{eq:lindblad_d_matrix}
\end{equation}
where the collision term 
$\frac{\partial \rho(t) }{\partial  t }\big|_{\mathrm{H}^{\rm col} }$
 accounts for the transitions between the eigenstates of $\mathrm{H}^{\rm har}$ driven by the perturbation $\mathrm{H}^{\rm col}$.
Here we are interested in the vibrational energy, which is a one-body operator at the harmonic leading order.
We can thus move to a formulation based on the one-body density matrix $\bm{\rho_1}$, which is defined in terms of a trace over the Fock space\cite{gebauer2004kinetic}: 
$\rho_{1}(\bm{q},\bm{q'},t)_{b\alpha,b'\hspace*{-0.5mm}\alpha'}{=}{\rm Tr} \big({\rho}(t)\;{a}^{\dagger}(\bm{q'})_{b'\hspace*{-0.5mm}\alpha'}\;{a}(\bm{q})_{b\alpha}\big)$.
It follows from equation~(\ref{eq:lindblad_d_matrix}) that $\bm{\rho_{1}}$ evolves according to:
\begin{equation}
\begin{split}
  &\frac{\partial }{\partial  t}{\bm{\rho_{1}}}(\bm{q},\bm{q'},t)+i\Big(\sqrt{\bm{\tenscomp{D}}(\bm{q})}\bm{\rho_{1}}(\bm{q},\bm{q'},t)-\bm{\rho_{1}}(\bm{q},\bm{q'},t)\sqrt{\bm{\tenscomp{D}}(\bm{q'})}\Big) \\
  &\hspace*{50mm}=\frac{\partial  }{\partial   t}{\bm{\rho_{1}}({\bm{q}},{\bm{q}'},t)}\Big|_{\mathrm{H}^{\rm col} }.
  \end{split}
  \raisetag{6mm}
  \label{eq:evol_density_matrix}
\end{equation}
For systems that are homogeneous in real space $\bm{\rho_{1}}$ is diagonal:  $\bm{\rho_{1}}(\bm{q},\bm{q'},t){=}\bm{\rho_{1}}(\bm{q},\bm{q'},t)\delta_{\bm{q},\bm{q'}}$;
to study non-homogeneous systems 
it becomes useful to introduce the Wigner distribution $\tens{W}$, obtained by applying the Wigner-Weyl transformation to $\bm{\rho_{1}}$\cite{frensley1990boundary}:
\begin{equation}
\begin{split}
  {\tenscomp{W}}{(}{\bm{R}},{\bm{q}}\hspace*{-0.3mm},{t}{)}_{\hspace*{-0.4mm}{b}{\alpha}{,}{b'}\hspace*{-0.5mm}{\alpha'}}{=}{\sum\limits_{\bm{q''}}} {\rho_{1}}{\big(}{\bm{q}}{+}{\bm{q''}}\hspace*{-0.7mm},{\bm{q}}{-}{\bm{q''}}\hspace*{-0.7mm},{t}{\big)}_{\hspace*{-0.5mm}{b}{\alpha}{,}{b'}\hspace*{-0.5mm}{\alpha'}}e^{2i\bm{q}''{\cdot} \bm{R}}\hspace*{-0.5mm}.\hspace*{7mm}
  \raisetag{8mm}
  \end{split}
  \label{eq:Wigner_transf_3D_momentum}
\end{equation}
It becomes apparent from equation~(\ref{eq:Wigner_transf_3D_momentum}) that a non-homogeneity in real space 
is related to a one-body density matrix with off-diagonal elements in reciprocal space.  
Now, we consider the out-of-equilibrium regime of a system perturbed by a small temperature gradient, appreciable over a length scale $L$ much larger than the norms of the direct lattice vectors $|\bm{a}_i|$ ($i{=}1,2,3$).  
Applying the transformation~(\ref{eq:Wigner_transf_3D_momentum}) to equation~(\ref{eq:evol_density_matrix}), we can obtain the evolution for the Wigner distribution. 
Such evolution equation can be greatly simplified by using the hypothesis of having non-homogeneities only at mesoscopic scales or above, \textit{i.e.} of having a one-body density matrix $\bm{\rho_{1}}{\big(}{\bm{q}}{+}{\bm{q''}}\hspace*{-0.7mm},{\bm{q}}{-}{\bm{q''}}\hspace*{-0.7mm},{t}{\big)}$ that is sharply peaked and significantly different from zero only for $|\bm{q''}|{\ll}{2\pi}|\bm{a}_{i}|^{-1}$.
This allows us to perform a Taylor expansion for $\sqrt{\tens{D}\big({\bm{q}}{\pm}\bm{q''}\big)}$ around $\bm{q''}{\to}0$, 
obtaining the  simplified equation:
\begin{equation}
\abovedisplayskip=2mm
\belowdisplayskip=2mm
\begin{split}
  &\frac{\partial  }{\partial  t} \tens{W}({\bm{R}},{\bm{q}},t)
+i\Big[\sqrt{\bm{\tenscomp{D}}(\bm{q})},\tens{W}(\bm{R},\bm{q},t)\Big]+\hspace*{20mm}\\
&+\frac{1}{2}\Big\{\vec{\nabla}_{\bm{q}} \sqrt{{\bm{D}}(\bm{q})},\cdot\vec{\nabla}_{\bm{R}} \tens{W}(\bm{R},\bm{q},t)\Big\}=\frac{\partial  }{\partial   t} \tens{W}(\bm{R},{\bm{q}},t)  \Big|_{\mathrm{H}^{\rm col} };
\label{eq:Wigner_evolution_equation}
\end{split}
\raisetag{14mm}
\end{equation}
where the gradients appearing in the anticommutator are contracted using the scalar product.
The spatial gradient $\vec{\nabla}_{\bm{R}}$ acts on the variable $\bm{R}$ appearing in the exponential $e^{2i \bm{q''}\cdot \bm{R}}$, whose characteristic length scale is $L{\sim}|\bm{q}''|^{-1}{\gg} |\bm{a}_{i}|$. Over such mesoscopic scale $L$ the discrete variable $\bm{R}$ can be considered as continuous and the gradient $\vec{\nabla}_{\bm{R}}$ well-defined.
At this point, thanks to the locality in $\bm{q}$ of equation~(\ref{eq:Wigner_evolution_equation}), we can recast it in the phonon representation by applying the unitary transformation that diagonalizes ${\tens{D}(\bm{q})}$:
$\mathcal{E}^{*}(\bm{q})_{{s},b\alpha}{\tenscomp{D}(\bm{q})}_{b\alpha,b'\hspace*{-0.5mm}\alpha'} \mathcal{E}(\bm{q})_{s'\hspace*{-0.5mm},b'\hspace*{-0.5mm}\alpha'}{=}\omega^2(\bm{q})_s\delta_{s,s'}$. We thus introduce the following quantities
\begin{align}
\abovedisplayskip=2mm
\belowdisplayskip=2mm
   &\tenscomp{N}(\bm{R},\bm{q},t)_{s,s'}{=}\mathcal{E}^{*}(\bm{q})_{{s},b\alpha}{\tenscomp{W}(\bm{R},\bm{q},t)}_{b\alpha,b'\hspace*{-0.5mm}\alpha'} \mathcal{E}(\bm{q})_{s'\hspace*{-0.5mm},b'\hspace*{-0.5mm}\alpha'},\hspace*{3mm} \label{eq:Population_Peierls_Generalized}\raisetag{10mm}\\[1mm]
 &\tenscomp{V}^\beta(\bm{q})_{s,s'}{=}\mathcal{E}^{*}(\bm{q})_{{s},b\alpha}{{\nabla^{\beta}_{\bm{q}} \sqrt{\tens{D}(\bm{q})}} }_{b\alpha,b'\hspace*{-0.5mm}\alpha'} \mathcal{E}(\bm{q})_{s'\hspace*{-0.5mm},b'\hspace*{-0.5mm}\alpha'}\hspace*{5mm}
 \label{eq:vel_op}
\end{align}
where the atomic and Cartesian indexes are replaced by the phonon branch index $s$.
We define the diagonal matrix $\Omega(\bm{q})_{s,s'}{=}\omega(\bm{q})_s\delta_{s,s'}$ 
and denote with $\frac{\partial \tenscomp{N}(\bm{R},{\bm{q}},t)_{s,s'} }{\partial   t}   \big|_{\mathrm{H}^{\rm col} }$ the 
collision operator in the basis of the phonon modes.
Equations~(\ref{eq:Population_Peierls_Generalized}) and (\ref{eq:vel_op}) yield respectively a generalization of the phonon distributions 
and group velocities appearing in the BTE\cite{peierls1929kinetischen,fugallo2013ab}; 
in the absence of degeneracies (\textit{i.e.} for $\omega(\bm{q})_s{\neq}\omega(\bm{q})_{s'}\forall s{\neq} s'$) the diagonal elements $s{=}s'$ of equation~(\ref{eq:Population_Peierls_Generalized}) and of equation~(\ref{eq:vel_op}) coincide, respectively, with the phonon distributions appearing in Peierls'  equation and with their group velocities. 
Equation~(\ref{eq:Wigner_evolution_equation}) in the phonon representation reads
\begin{widetext}
\begin{equation}
\frac{\partial  }{\partial  t} \tens{N}({\bm{R}},{\bm{q}},t)
+i\Big[\bm{\Omega}(\bm{q}),\tens{N}(\bm{R},\bm{q},t)\Big]+\frac{1}{2}\Big\{\vec{\tens{V}}(\bm{q}),\cdot \vec{\nabla}_{\bm{R}} \tens{N}(\bm{R},\bm{q},t)\Big\}=\frac{\partial }{\partial  t}\tens{N}(\bm{R},{\bm{q}},t)   \bigg|_{\mathrm{H}^{\rm col} }.\label{eq:Wigner_evolution_equation_N}
\end{equation}
\end{widetext}
This equation is the first main result of this paper, and offers a more general phonon transport  equation, where phonon wave packets are not only allowed to propagate particle-like in space but also to tunnel, wave-like, from one branch to another.  
We note that the left-hand side of equation~(\ref{eq:Wigner_evolution_equation_N}) does not contain $\hbar$, and it does not distinguish between fermions or bosons. Quantum statistics is instead enforced by the collision term at the right-hand side (see e.g. equation~(\ref{eq:scattering_operator}) later).
The vibrational energy density can now be determined 
from the solution of equation~(\ref{eq:Wigner_evolution_equation_N}): $E(\bm{R},t){=}\frac{\hbar}{\mathcal{V}N_{\rm c}}\sum_{\bm{q},s}\big(\tens{N}(\bm{R},\bm{q},t)\bm{\Omega}(\bm{q})\big)_{s,s}$.
Computing the time derivative of $E(\bm{R},t)$ and using equation~(\ref{eq:Wigner_evolution_equation_N}), one obtains the harmonic heat flux $\bm{J}(\bm{R},t)$:
\begin{equation}
    \bm{J}(\bm{R},t){=}
  \frac{1}{2}\frac{1}{\mathcal{V}N_{\rm c}}\sum_{\bm{q},s} {\Big(\big\{ \vec{\tens{V}}(\bm{q}), \tens{N}(\bm{R},\bm{q},t)\big\}\hbar \bm{\Omega}(\bm{q}) \Big)}_{s,s}.
\label{eq:heat_flux}
\raisetag{8mm}
\end{equation}
To compute the thermal conductivity we focus on the steady-state regime and search for a solution of equation~(\ref{eq:Wigner_evolution_equation_N}) linear in the temperature gradient. 
The right-hand side collision term 
can be written in a closed form approximation expressing the many-body collision operator~(\ref{eq:lindblad_d_matrix}) in terms of one-body quantities and linearizing it around the equilibrium temperature $T$\cite{frensley1990boundary,gebauer2004kinetic,cohen2004atom} 
\begin{equation}
\medmuskip=0mu
\thinmuskip=0mu
\thickmuskip=0mu
\begin{split}
  \frac{\partial }{\partial   t} &\tenscomp{N}({\bm{R},\bm{q},t})_{\hspace*{-0.5mm}s{,}s'}    \hspace*{-0.5mm}\Big|_{\mathrm{H}^{\rm col} }\hspace*{-0mm}{=}\hspace*{1mm}-(1-\delta_{s,s'})\frac{\Gamma(\bm{q})_{s}+\Gamma(\bm{q})_{s'}}{2} \tenscomp{N}(\bm{R},\bm{q},t)_{s,s'}\\
 &{-}\frac{\delta_{s,s'}}{\mathcal{V}N_{\rm c}}\hspace*{-0.5mm}{\sum\limits_{{s''}{\bm{q}''}}}\hspace*{-0.5mm} \tenscomp{{A}}^T(\bm{q},{\bm{q}''})_{\hspace*{-0.4mm}s,{s''}}\hspace*{-0.5mm}\big(\hspace*{-0.5mm}\tenscomp{N}(\bm{R},{\bm{q}''}\hspace*{-1mm},t)_{\hspace*{-0.4mm}{s''}\hspace*{-0.5mm},{s''}}{-}\bar{\tenscomp{N}}^T\hspace*{-1mm}({\bm{q}''})_{\hspace*{-0.4mm}{s''}}\hspace*{-0.5mm}\big);
  \raisetag{8mm}\label{eq:scattering_operator}
\end{split}
\end{equation}
where $\mathcal{V}$ is the unit cell volume,
$\bar{\tenscomp{N}}^T\hspace*{-1mm}({\bm{q}})_{\hspace*{-0.2mm}{s}}{=}\big({\exp\hspace*{-1mm}\big[\hspace*{-0.5mm}\frac{\hbar\omega(\hspace*{-0.2mm}\bm{q}\hspace*{-0.2mm})_{\hspace*{-0.2mm}{s}}}{k_B T} \hspace*{-0.5mm}\big]{-}1}\big)^{-1}\hspace*{-1mm}$ is the equilibrium Bose-Einstein distribution, $\Gamma(\bm{q})_{s}{=}(\mathcal{V}N_{\rm c})^{-1}\tenscomp{A}^T(\bm{q},{\bm{q}})_{s,s''}\delta_{s'',s}$ is the phonon linewidth and $\tenscomp{A}^T(\bm{q},{\bm{q}''})_{s,{s''}}$ is the standard collision operator that accounts for anharmonicity and isotopic disorder 
(for the third-order anharmonicities considered here, 
see equation~(13) of Ref.\cite{fugallo2013ab}).
Under these conditions, equation~(\ref{eq:Wigner_evolution_equation_N}) can be decoupled in its diagonal and off-diagonal parts, as shown in Methods;
these two equations describe the evolution of the populations and coherences, respectively, using the standard nomenclature in this field.
The equation for the populations turns out to be the standard Peierls' LBTE that can be solved exactly\cite{fugallo2013ab,omini1995iterative,carrete2017almabte,PhysRevLett.110.265506,cepellotti2016thermal};
the equation for the coherences can be solved straightforwardly (see Methods). 
Once the solution $\tenscomp{N}(\bm{R},\bm{q})_{s,s'}$ for both populations ($s{=}s'$) and coherences ($s{\neq} s'$)  
is known,  one can compute the heat flux ${J}^\alpha(\bm{R})$ and  the thermal conductivity $\kappa^{\alpha\beta}$ from
${J}^\alpha(\bm{R}){=}{-}\kappa^{\alpha\beta}\nabla T^\beta(\bm{R})$. 
The second main result of this work follows, where the thermal conductivity is generalized to an expression including both the populations' and coherences' contributions: 
\begin{widetext}
\begin{equation}
\begin{split}
\kappa^{\alpha \beta}=\kappa^{\alpha \beta}_{\rm P}+\frac{\hbar^2}{k_{B} {T}^2}\frac{1}{\mathcal{V}N_{\rm c}}&\sum_{\bm{q}}\sum_{s\neq s'}\frac{\omega(\bm{q})_{s}+\omega(\bm{q})_{s'}}{2}{\tenscomp{V}^\alpha}(\bm{q})_{s,s'}{\tenscomp{V}}^\beta(\bm{q})_{s',s}\times\\
&\times\frac{\omega(\bm{q})_{s}\bar{\tenscomp{N}}^{T}({\bm{q}})_{s}[\bar{\tenscomp{N}}^{T}({\bm{q}})_{s}+1]+\omega(\bm{q})_{s'}\bar{\tenscomp{N}}^{T}({\bm{q}})_{s'}[\bar{\tenscomp{N}}^{T}({\bm{q}})_{s'}+1]}{4[\omega(\bm{q})_{s'}-\omega(\bm{q})_{s}]^2+[\Gamma(\bm{q})_{s}+\Gamma(\bm{q})_{s'}]^2}[\Gamma(\bm{q})_{s}+\Gamma(\bm{q})_{s'}];\label{eq:thermal_conductivity_final_sum}
\end{split}
\end{equation}
\end{widetext}
where $\kappa^{\alpha\beta}_{\rm P}$ is the standard Peierls contribution to the conductivity  and the additional tensor, which we name $\kappa^{\alpha\beta}_{\rm C}$, derives from the coherences' equation.
The relevance of equation~(\ref{eq:thermal_conductivity_final_sum}) is its capability to account for heat transfer associated to both the 
diagonal (populations) and off-diagonal (coherences) Wigner distribution elements. The former is associated to the particle-like propagation of phonon wave packets discussed by Peierls'  semiclassical picture; the latter is related to the wave-like tunneling and loss of 
coherence between different vibrational eigenstates.
Equation~(\ref{eq:thermal_conductivity_final_sum}) highlights the conditions under which $\kappa^{\alpha\beta}_{\rm C}$ is relevant, namely when the difference between the phonon frequencies is of the same order of the average phonon linewidth and, at the same time, 
the velocity operator has non-negligible off-diagonal elements --- we will show later that this is what precisely happens in complex crystals. 
In simple crystals, instead, the present approach and the standard LBTE provide very similar results since, as we will show 
later,
$\kappa^{\alpha\beta}_{\rm P}{\gg}\kappa^{\alpha\beta}_{\rm C}$. 
Since the thermal conductivity~(\ref{eq:thermal_conductivity_final_sum}) derives from an exact linear-order solution of equation~(\ref{eq:Wigner_evolution_equation_N}), it remains valid also in the hydrodynamic regime\cite{cepellotti2015phonon}. In the kinetic regime\cite{cepellotti2015phonon} instead, 
a good estimation of $\kappa^{\alpha\beta}_{\rm P}$ can be obtained using the SMA approximation, \textit{i.e.} neglecting the scattering events that repopulate the diagonal $\tenscomp{N}(\bm{R},{\bm{q}},t)_{{s},{s}}$\cite{fugallo2013ab}. 
In practice, the SMA consists in replacing the collision matrix for the populations of equation~(\ref{eq:scattering_operator}) with  its diagonal elements only: 
$\tenscomp{A}^T(\bm{q},{\bm{q''}})_{s,s''}{\simeq}(\mathcal{V}N_{\rm c})\Gamma(\bm{q})_{s}\delta_{\bm{q},\bm{q''}}\delta_{s,s''}$. 
The equation for the coherences is unaffected by this approximation\cite{cohen2004atom};
it follows that the total SMA conductivity is given by an expression equal to the last term of equation~(\ref{eq:thermal_conductivity_final_sum}), but with the summation running in full over $s$ and $s'$ (equation~(\ref{eq:thermal_conductivity_final_sum_symmetric}) in Methods).  
Last, we show in Methods how in the limit of a harmonic crystal with an infinitely large unit cell (\textit{i.e.} with the Brillouin zone reduced to the $\Gamma$-point) the normalized trace of the coherences' conductivity tensor $\frac{1}{3}\kappa^{\alpha\alpha}_{\rm C}$ reduces to the Allen-Feldman expression for glasses\cite{allen1989thermal}. 

The perovskite CsPbBr$_3$, which belongs to a family of materials promising for thermoelectric energy conversion\cite{wang2018cation,lee2017ultralow}, is a good case study for this more general framework, as it features many quasi-degenerate phonon branches and its thermal properties\cite{wang2018cation,lee2017ultralow} are not correctly described by the LBTE\cite{lee2017ultralow}.
We compute from first-principles the harmonic Hamiltonian and the collision operator (see details in Methods), calculating the complete thermal conductivity of equation~(\ref{eq:thermal_conductivity_final_sum}) as a function of temperature.
\begin{figure}[b!]
\vspace*{-4mm}
  \centering
  \includegraphics[width=\columnwidth]{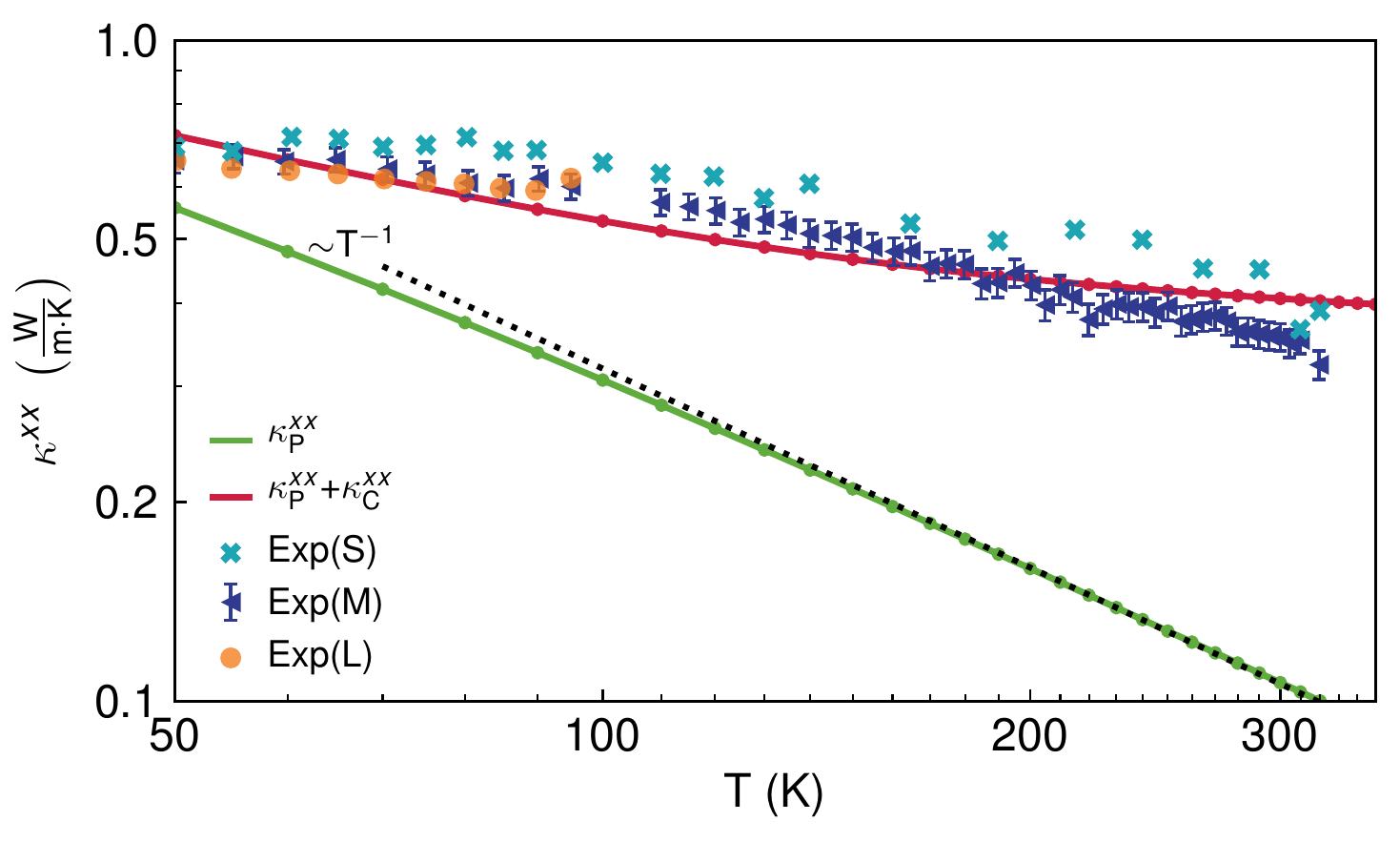}
   \vspace{-6mm}
  \caption{\sf\textbf{Bulk thermal conductivity of CsPbBr$_3$ as a function of temperature.} 
    ``Exp(L)'', ``Exp(M)'' and ``Exp(S)''  refer to experiments\cite{wang2018cation} on nanowires having respectively sections of $800{\times} 380\;\rm{nm}^2$, $320{\times} 390\;\rm{nm}^2$ and $300{\times} 160\;\rm{nm}^2$: their broad agreement supports the hypothesis of negligible finite-size boundary scattering\cite{wang2018cation}. Green, Peierls' LBTE conductivity, with its universal $T^{-1}$ asymptotics (dotted line)\cite{sun2010lattice,lory2017direct,li2015ultralow,ziman1960electrons}. Red, conductivity from equation~(\ref{eq:thermal_conductivity_final_sum}).
  }
  \label{fig:k_vs_T}
\end{figure}
\begin{figure*}[t!]
\vspace*{-8mm}
  \centering
  \begin{overpic}[width=0.6\textwidth]{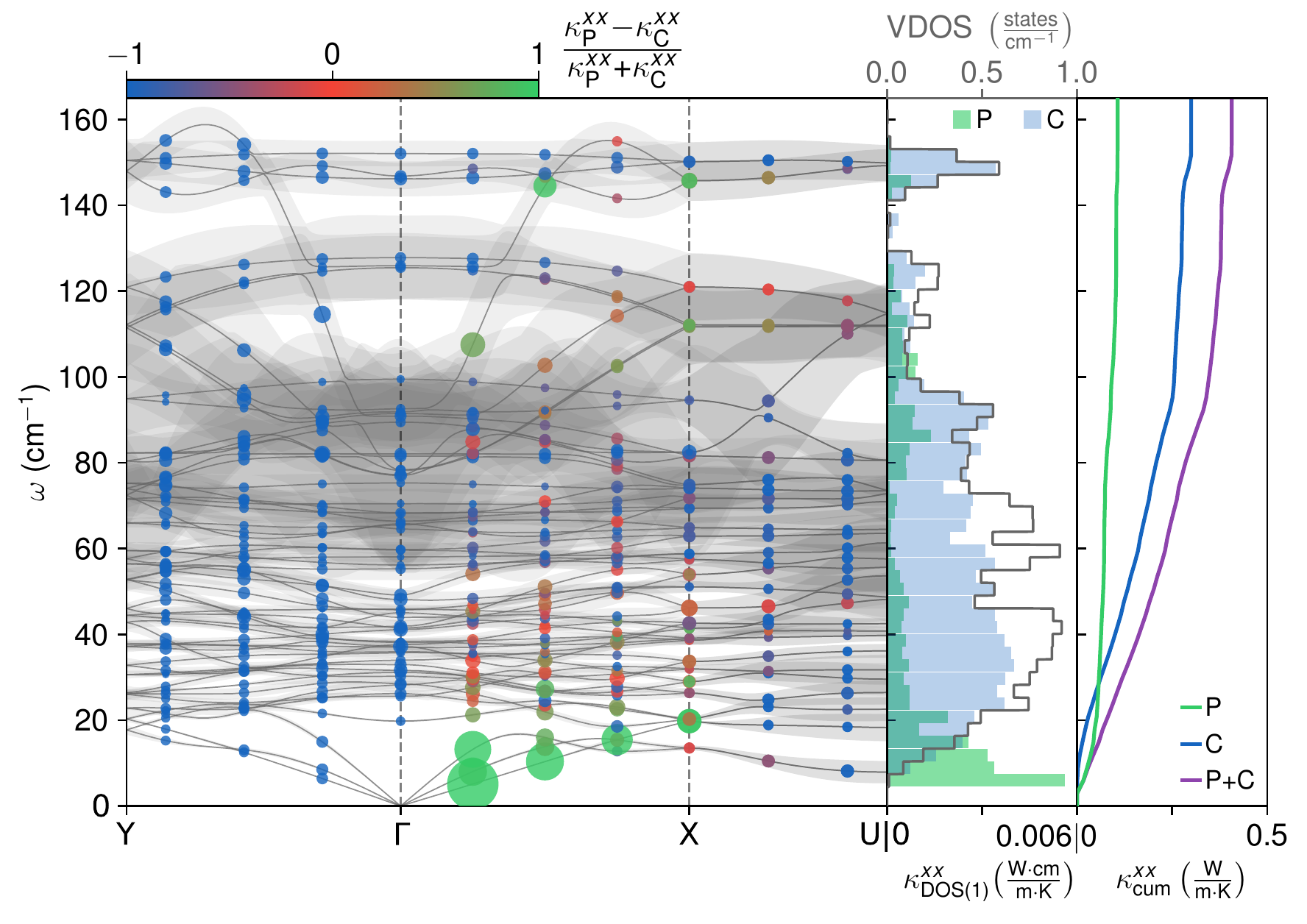}
  \put (10,60.7) {\sf\textbf{a}}
  \put (69,60.7) {\sf\textbf{b}}
  \put (83.5,60.7) {\sf\textbf{c}}
  \end{overpic}
  \begin{minipage}[b]{0.39\textwidth}
    \begin{overpic}[width=0.9\textwidth]{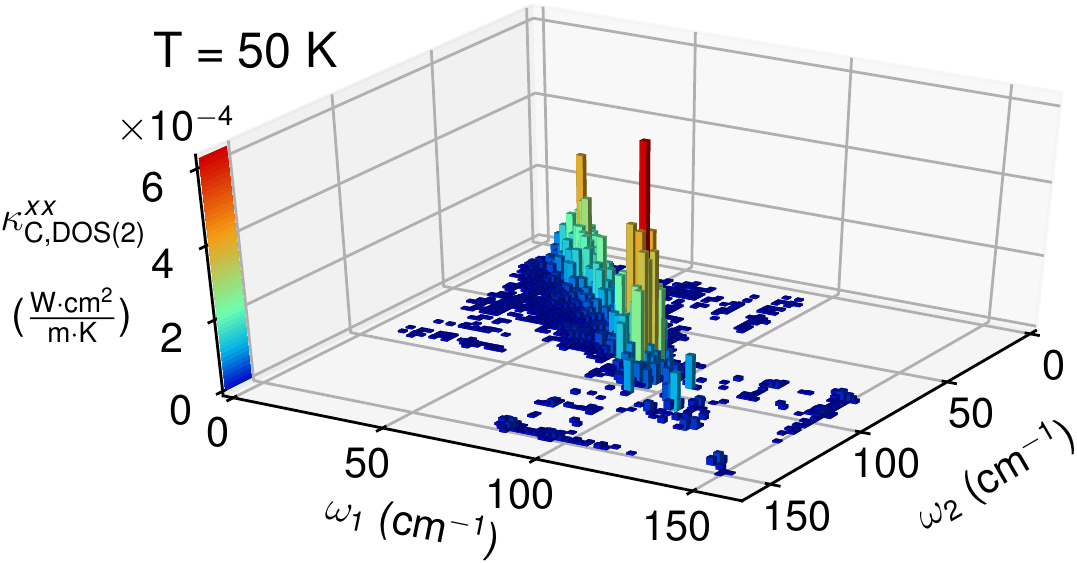}
    \put (0,46.5) {\sf\textbf{d}}
    \end{overpic}
    \vspace*{3.5mm}
    \begin{overpic}[width=0.9\textwidth]{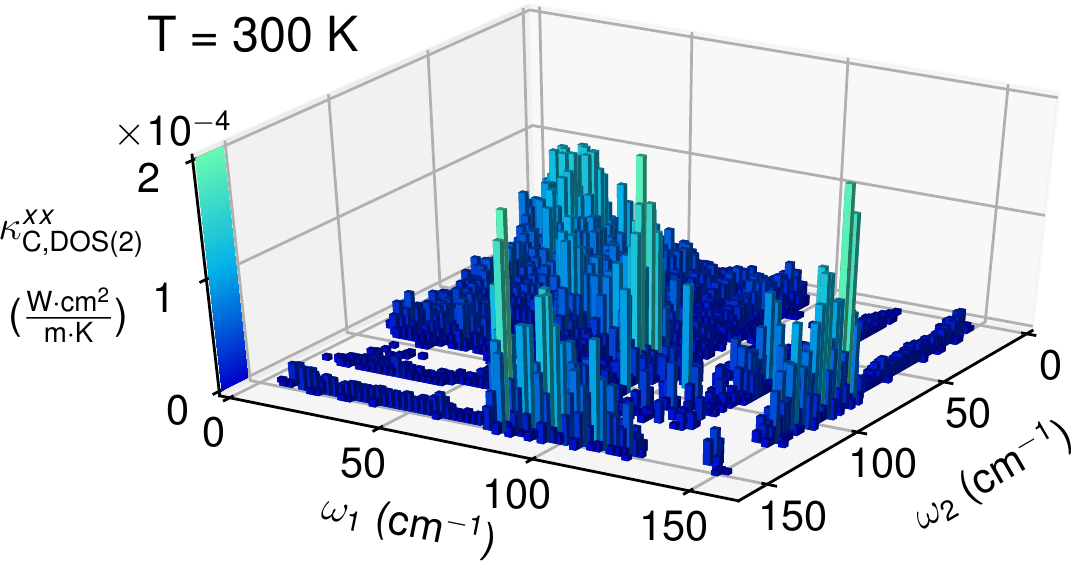}
     \put (0,48.2) {\sf\textbf{e}}
    \end{overpic}
    \vspace*{2mm}
  \end{minipage}\\[-3mm]
\caption{\sf\textbf{Vibrational properties and heat conduction mechanisms in CsPbBr$_3$.} \textbf{a},~The phonon spectrum of CsPbBr$_3$ (gray lines) features low-energy flat branches, typical of rattling vibrations\cite{voneshen2013suppression,lee2017ultralow,li2015ultralow}. The shaded gray areas are half the phonon linewidths ($\Gamma(\bm{q})_s/2$ for graphical clarity) at 300~K. 
Colored circles in the plot represent the  phonon eigenstates $(\bm{q},s)$ sampled in the  calculation; the area of each circle is related to its contribution to $\kappa^{xx}$ at 300~K and the color identifies the origin of the contribution: green for populations' propagation and blue for coherences' anharmonic couplings (red corresponds to $50\%$ of each). In the coherences' couplings between two modes $(\bm{q},s)$ and $(\bm{q},{s'})$ the contribution of the single mode $s$ is determined as ${C(\bm{q})_s}/[{C(\bm{q})_s{+}C(\bm{q})_{s'}}]$,  
 where $C(\bm{q})_s{=}{\hbar^2}/(k_{B} {T}^2)\omega^2\hspace*{-0.5mm}(\bm{q})\hspace*{-0.4mm}_{s}\bar{\tenscomp{N}}^{T}\hspace*{-0.6mm}({\bm{q}})_{\hspace*{-0.4mm}s}[\bar{\tenscomp{N}}^{T}\hspace*{-0.5mm}({\bm{q}})_{\hspace*{-0.4mm}s}{+}1]$ 
 is the specific heat of a given phonon.
\textbf{b},~Thermal conductivity density of states $\kappa^{xx}_{\rm{DOS(1)}}$ of populations (``P'', green) and coherences (``C'', blue). The gray line shows the vibrational density of states (VDOS). \textbf{c},~Cumulative thermal conductivity (in purple) as a sum of the populations' contribution (``P'') in green and 
coherences' one (``C'') in blue. \textbf{d},~\textbf{e},~2-dimensional density of states $\kappa^{xx}_{\rm{C,DOS}(2)}$ for the contribution to the thermal conductivity of the coherences term at 50~K (\textbf{d}) and 300~K (\textbf{e}). The points on the diagonal $\omega_1{=}\omega_2$ correspond to degenerate eigenstates; the Allen-Feldman framework considers couplings only between these.} 
  \label{fig:cond_mech}
  \vspace*{-3mm}
\end{figure*}
The results are shown in figure~(\ref{fig:k_vs_T}) (we show here the SMA for $\kappa^{xx}$; this is almost identical to the full solution (see Methods)): at 300~K  the populations term contributes just $30\%$ of the total conductivity, while the 
coherences term provides an additional $70\%$, leading to much closer agreement with experiments\cite{wang2018cation} that becomes even more relevant in the high-temperature asymptotics.
Conversely, at low temperature Peierls'  term becomes dominant (at 50~K it provides 78$\%$ of the total conductivity). 
It can be seen that in the high temperature limit $\kappa^{\alpha\beta}_{\rm P}{\propto}\;T^{-1}$, as  predicted by Peierls'  theory\cite{sun2010lattice,lory2017direct,li2015ultralow,ziman1960electrons}; this in broad disagreement with experiments. Instead, the present theory predicts a decay of $\kappa^{xx}$ much milder than $T^{-1}$ as shown here for CsPbBr$_3$, and as also present in many other complex crystals\cite{li2015ultralow,lory2017direct,PhysRevB.46.6131,PhysRevB.96.214202,chen2015twisting,voneshen2013suppression}. 
We note in passing that the $\kappa^{\alpha\beta}_{\rm P}$ calculated here and the one presented in Ref.\cite{lee2017ultralow} are very similar, confirming that all computational parameters are correctly accounted for. 
\mbox{Figure~(\ref{fig:cond_mech}\textbf{a-c})} shows that phonon branches with large group velocities and low anharmonicity provide a large contribution to the populations term; conversely, highly anharmonic flat bands contribute to the coherences term.
The populations contribution originates mainly from low-frequency modes, while all the spectrum contributes to the 
coherences term. 
At 50~K, the 2-dimensional  density of states for the coherences' thermal conductivity ($\kappa^{xx}_{\rm{C,DOS}(2)}$, figure~(\ref{fig:cond_mech}\textbf{d}))  shows couplings between quasi-degenerate states similarly to the case of harmonic  glasses ($\Gamma(\bm{q})_s{\to}0$). 
At 300~K (figure~(\ref{fig:cond_mech}\textbf{e})) $\kappa^{xx}_{\rm{C,DOS}(2)}$ instead includes couplings between phonon modes having very different frequencies, driven by the large anharmonicity ---  therefore the corresponding heat conduction mechanism is intrinsically 
different from the one of harmonic glasses.
In order to verify that the present theory reduces to the Peierls regime in the limit of simple crystals, we computed the room-temperature (300~K) thermal conductivity~(\ref{eq:thermal_conductivity_final_sum}) of silicon and diamond.  
We found that for these crystals the coherences term is negligible: for silicon $\kappa_{\rm P}^{xx}{=}142.83\;\rm{W}/(\rm{m}{\cdot}\rm{K})$ vs. $\kappa_{\rm C}^{xx}{=}0.23\;\rm{W}/(\rm{m}{\cdot}\rm{K})$;  for diamond $\kappa_{\rm P}^{xx}{=}2780.07\;\rm{W}/(\rm{m}{\cdot}\rm{K})$ vs. $\kappa_{\rm C}^{xx}{=}0.03\;\rm{W}/(\rm{m}{\cdot}\rm{K})$.

In summary, we have derived from the Wigner phase space formulation of quantum mechanics a general transport equation~(\ref{eq:Wigner_evolution_equation_N}) 
able to account rigorously and simultaneously for anharmonicity and disorder, leading to a thermal conductivity~(\ref{eq:thermal_conductivity_final_sum}) that adds an interbranch tunneling term for the coherences;
this equation~(\ref{eq:thermal_conductivity_final_sum}) reduces to the Peierls' BTE in the limit of simple crystals, and it provides a rigorous  derivation of the BTE for phonon populations first heuristically derived  by Peierls\cite{peierls1929kinetischen}.
Also, in the limit of harmonic glasses, the thermal conductivity~(\ref{eq:thermal_conductivity_final_sum}) reduces to the Allen-Feldman limit; equation~(\ref{eq:thermal_conductivity_final_sum}) is more general and encompasses all intermediate cases, applying also to anharmonic glasses.
The LBTE approach, which accounts for populations only, greatly underestimates the thermal conductivity of complex crystals\cite{li2015ultralow,lee2017ultralow,chen2015twisting,PhysRevB.96.214202,lory2017direct}
that ubiquitously display a glass-like high-temperature behavior (\textit{i.e.} decay milder than the universal $T^{-1}$ trend predicted by Peierls'  theory\cite{sun2010lattice,lory2017direct,li2015ultralow,ziman1960electrons}) and that up to now has been only heuristically explained by several models\cite{PhysRevB.46.6131,agne2018minimum,Clarke2003,chen2015twisting}.
In the present approach, the correct conductivity is recovered, and its high-temperature decay (figure~(\ref{fig:k_vs_T})).

We note in passing that a complementary approach to these transport equations is to compute the thermal conductivity via the
 Green-Kubo method\cite{marcolongo2016microscopic,seyf2017rethinking,lv2016non,carbogno2017ab,puligheddu2017first},
which is based on molecular dynamics and is exact, although computationally more expensive, provided one is in the classical Maxwell-Boltzmann high-temperature limit (\textit{i.e.} at $T{\gg} T_D$, where $T_D$ is the Debye temperature).
Coherences' effects have been thoroughly studied in the context of electronic transport\cite{gebauer2004kinetic,PhysRevB.86.155433,frensley1990boundary}
 and it is worth mentioning that
the term $\kappa^{\alpha\beta}_{\rm C}$ has some analogies with the electronic current related to the Zener interband transitions\cite{PhysRevB.86.155433}. 
However, thermal transport is fundamentally different from electronic transport since the full phonon spectrum contributes to the heat flow, while for electrons the current originates only from the bands close to the Fermi level.
This novel formulation is also going to be relevant to applied research, as it allows to predict the low thermal conductivity 
of target materials for thermoelectric energy conversion\cite{lee2017ultralow,voneshen2013suppression}.

\noindent
{\sf{\textbf{\large{Acknowledgements}}}}\\

\noindent
{\sf{\textbf{\large{Author contributions}}}}\\
The project was conceived by all authors. 
M.S. performed the numerical calculations and prepared
the figures with inputs from N.M and F.M. 
All authors contributed to the redaction of the manuscript. \\

\noindent
{\sf{\textbf{\large{Competing interests}}}}\\
The authors declare no competing interests.\\

\noindent
{\sf{\textbf{\large{Additional information}}}}\\
\textbf{Correspondence and request of materials}  should be addressed to F.M.
\vspace*{5mm}

\noindent
{\sf{\textbf{\large{Methods}}}}\\
\textbf{Bosonic operators in the Cartesian basis.} 
In order to describe heat transfer driven by a temperature gradient (\textit{i.e.} a spatially-dependent lattice vibrational energy), we need to keep track of the vibrations' location in real space.
This is possible using the Fourier transform of the bosonic operator defined in equation~(\ref{eq:def_op_a}):
\begin{equation}\label{eq:def_op_a_real}
  \begin{split}
  \mathrm{a}(\bm{R})_{b\alpha}{=}&\frac{1}{\sqrt{N_{\rm c}}}\sum_{\bm{q}}\mathrm{a}(\bm{q})_{b\alpha}e^{+ i \bm{q}\cdot \bm{R}}{=}\\
  {=}&\frac{1}{\sqrt{2 \hbar}}\sum_{\bm{R''}} \bigg(\sqrt[4]{\tenscomp{G}^{{-}1}}_{\bm{R}b\alpha,(\bm{R}{+}\bm{R''})b'\alpha'}\frac{{\mathrm{p}}(\bm{R}{+}\bm{R''})_{b'\alpha'}}{\sqrt{M_{b'}}} \\&{-}i \sqrt{M_{b'}}{\mathrm{u}}(\bm{R}{+}\bm{R''})_{b'\alpha'} \sqrt[4]{\tenscomp{G}}_{(\bm{R}{+}\bm{R''})b'\alpha'\hspace*{-0.5mm},\bm{R}b\alpha}\bigg);
  \end{split}
  \raisetag{29mm}
\end{equation}
since, as we will show now, it is related to vibrations localized around the lattice site indexed by $(\bm{R},b,\alpha)$. 
The tensor ${\tenscomp{G}}_{\bm{R}b\alpha,{(\bm{R}{+}\bm{R''})}b'\alpha'}$ appearing in equation~(\ref{eq:def_op_a_real}) is related to the interatomic harmonic force constants in real space\cite{ziman1960electrons}, $\Phi_{\bm{R}b\alpha,(\bm{R}{+}\bm{R''})b'\alpha'}$:
\begin{equation}
 {\tenscomp{G}}_{\bm{R}b\alpha,{(\bm{R}{+}\bm{R''})}b'\alpha'} {=}\frac{\Phi_{\bm{R}b\alpha,(\bm{R}{+}\bm{R''})b'\alpha'}}{\sqrt{M_b M_{b'}}};
\end{equation}
it is translation-invariant and decays to zero for $\bm{R''}{\to}\infty$\cite{ziman1960electrons,Simoncelli_full_2018}. 
The harmonic Hamiltonian~(\ref{eq:full_H}) can be written in terms of the vibrational energy field $\mathrm{H}^{\rm har}{=}\sum\limits_{\bm{R},b,\alpha} \mathcal{H}(\bm{R})_{b,\alpha}$, where
\begin{equation}
\begin{split}
    &\mathcal{H}(\bm{R})_{b\alpha}{=}{\frac{\mathrm{p}^2(\bm{R})_{b\alpha}}{2 {M_{b}}}} +\\
    &+\frac{1}{2} \bigg(\sum_{\scriptscriptstyle{\bm{R'}} } {\sqrt{\tenscomp{G}}_{\bm{R}b\alpha,(\bm{R}{+}\bm{R'})b'\hspace*{-0.5mm}\alpha'}}\sqrt{M_{b'}}\mathrm{u}\left(\bm{R}{+}\bm{R'}\right)_{b'\hspace*{-0.5mm}\alpha'}\bigg){\times}\\
    &
    {\times}\bigg(\sum_{\scriptscriptstyle{\bm{R''}} } {\sqrt{\tenscomp{G}}_{\bm{R}b\alpha,(\bm{R}{+}\bm{R''})b''\hspace*{-0.5mm}\alpha''}}\sqrt{M_{b''}}\mathrm{u}\left(\bm{R}{+}\bm{R''}\right)_{b''\hspace*{-0.5mm}\alpha''}\bigg).
\end{split}
\label{eq:energy_field}
\raisetag{30mm}
\end{equation}
In terms of the bosonic operators $\mathrm{a}^{\dagger}(\bm{R})_{b\alpha},\;\mathrm{a}(\bm{R})_{b\alpha}$, the energy field~(\ref{eq:energy_field}) becomes: 
\begin{equation}
\begin{split}
    &\mathcal{H}(\bm{R})_{b\alpha}{=} \hbar \sum_{\bm{R'},\bm{R''}} 
    \sqrt[4]{\tenscomp{G}}_{\bm{R}b\alpha,(\bm{R}{+}\bm{R'})b'\hspace*{-0.5mm}\alpha'}
    \sqrt[4]{\tenscomp{G}}_{\bm{R}b\alpha,(\bm{R}{+}\bm{R''})b''\hspace*{-0.5mm}\alpha''}{\times}\\
    &{\times}\Big( 
           \mathrm{a}^{\dagger}(\bm{R}{+}\bm{R'})_{b'\hspace*{-0.5mm}\alpha'}
           \mathrm{a}(\bm{R}{+}\bm{R''})_{b''\hspace*{-0.5mm}\alpha''}
+\frac{1}{2}\delta_{\bm{R'},\bm{R''} }\delta_{b',b''}\delta_{\alpha',\alpha''}
           \Big).
\end{split}
\label{eq:energy_field_simple}
\end{equation}
From the decay to zero of ${\tenscomp{G}}_{\bm{R}b\alpha,{(\bm{R}{+}\bm{R''})}b'\alpha'}$ for $\bm{R''}{\to}\infty$, it follows that the excitation created by $\mathrm{a}^\dagger(\bm{R})_{b\alpha}$ has an energy distribution $E(\bm{R}{+}\bm{R''})_{{\bm{R}b\alpha}}$ localized around the position $\bm{R}$\cite{Simoncelli_full_2018}:
\begin{equation}
  \begin{split}
      E(\bm{R}{+}\bm{R''})_{{\bm{R}b\alpha}}{=}&\sum_{{b''\alpha''}}\big<0\big| \mathrm{a}(\bm{R})_{b\alpha}\mathcal{H}(\bm{R}{+}\bm{R''})_{b''\alpha''}\mathrm{a}^\dagger(\bm{R})_{b\alpha}\big|0\big>\\
      &-\sum_{{b''\alpha''}}\big<0\big| \mathcal{H}(\bm{R}{+}\bm{R''})_{b''\alpha''}\big|0\big>\\
      =&\hbar\sum_{{b''\alpha''}}\Big(\sqrt[4]{\tenscomp{G}}_{\bm{R''}b''\hspace{-0.5mm}\alpha'',{\bm{0}}{b}{\alpha}}\Big)^2, 
  \end{split}
  \raisetag{7mm}
  \label{eq:location_energy_4}
\end{equation}
where we have used the translational invariance of the tensor $\tens{G}$ to simplify the last equation: $\sqrt[4]{\tenscomp{G}}_{(\bm{R}{+}\bm{R''})b''\hspace{-0.5mm}\alpha'',{\bm{R}}{b}{\alpha}}{=}\sqrt[4]{\tenscomp{G}}_{\bm{R''}b''\hspace{-0.5mm}\alpha'',{\bm{0}}{b}{\alpha}}$.
We have therefore demonstrated that the operator~(\ref{eq:def_op_a_real}) creates vibrations localized around the lattice position $\bm{R}$; hence, performing the Fourier transform of the operator~(\ref{eq:def_op_a}) it is possible to keep track of the vibrations' locations in real space. 
In contrast, the more common phonon-eigenmodes basis, obtained by the transformation (discussed in equations~(\ref{eq:Population_Peierls_Generalized},\ref{eq:vel_op}) of the main text):
\begin{eqnarray}
  \mathrm{a}(\bm{q})_{s'}&=\mathcal{E}^{*}(\bm{q})_{{s'},b'\alpha'}\mathrm{a}(\bm{q})_{b'\alpha'};
  \label{eq:ph_op_mode_basis}
\end{eqnarray}
does not allow such a real space-based description. 
In fact, the operator $ \mathrm{a}(\bm{q})_{s'}$ does not have a direct correspondence with the location of vibrations in real space, due to the phase indeterminacy of the eigenstates $\mathcal{E}(\bm{q})_{{s'},b'\alpha'}$ of the dynamical matrix $\tens{D}(\bm{q})$. 
This becomes apparent considering the following gauge transformation, allowed by the phase freedom on $\mathcal{E}(\bm{q})_{{s'},b'\alpha'}$ at each $\bm{q}$-point:
\begin{equation}
  \tilde{\mathrm{a}}(\bm{q})_{s}=\mathrm{a}(\bm{q})_{s} e^{i\bm{q}{\cdot}\bm{R}_{\Delta}};
  \label{eq:shifted_operator_a}
\end{equation}
where $\bm{R}_{\Delta}$ is a direct lattice vector.
Despite $\tilde{\mathrm{a}}(\bm{q})_{s}$ and $\mathrm{a}(\bm{q})_{s}$ being equivalent for a description in Fourier space, 
their real space representations are shifted by $\bm{R}_{\Delta}$:
\begin{equation}
  \begin{split}
  &\mathrm{a}(\bm{R})_{s}{=}\frac{1}{\sqrt{N_{\rm c}}}\sum_{\bm{q}}\mathrm{a}(\bm{q})_{s}e^{+ i \bm{q}\cdot \bm{R}};\\
  & \tilde{\mathrm{a}}(\bm{R})_{s}{=}\frac{1}{\sqrt{N_{\rm c}}}\sum_{\bm{q}}\tilde{\mathrm{a}}(\bm{q})_{s}e^{+ i \bm{q}\cdot \bm{R}}{=}\mathrm{a}(\bm{R}{+}\bm{R}_{\Delta})_{s}.
  \label{eq:ill-definiteness}
  \end{split}
\end{equation}
It is worth mentioning that this mirrors the electronic case, where the phase indeterminacy of the Bloch orbitals is reflected in the non-uniqueness of the transformation into Wannier functions\cite{RevModPhys.84.1419}. 
In this work, the use of the Cartesian basis --- which has no phase freedom --- renders the real space description and hence the Wigner distribution~(\ref{eq:Wigner_transf_3D_momentum}) well-defined.
\\[2mm]

\noindent
\textbf{Decoupling of equation~(\ref{eq:Wigner_evolution_equation_N}).}
We consider the steady-state regime, in which a small, slowly varying temperature gradient $\nabla T$ is established throughout the system. We look for a solution linear in the temperature gradient $\nabla T$\cite{fugallo2013ab} on the distribution corresponding to the local equilibrium temperature\cite{hardy1970phonon} $T(\bm{x})$:
\begin{equation}
  \tenscomp{N}(\bm{x},{\bm{q}})_{s,s'}{=}\bar{\tenscomp{N}}^{T}\hspace*{-1mm}({\bm{q}})_s\delta_{s,s'}{+}\bar{\tenscomp{n}}^{T}\hspace*{-1mm}({\bm{x}}, {\bm{q}})_{s}\delta_{s,s'}{+}\vec{\tenscomp{n}}^{(1)}\hspace*{-0.5mm}({\bm{q}})_{s,s'}{\cdot}\nabla T(\bm{x});
  \label{eq:expansion}
\end{equation}
where $\bar{\tenscomp{n}}^{T}({\bm{x}}, {\bm{q}})_{s}\delta_{s,s'}$ accounts for the local equilibrium temperature: $\bar{\tenscomp{n}}^{T}({\bm{x}}, {\bm{q}})_{s}{=}
\frac{d \bar{\tenscomp{N}}^{T}({\bm{q}})_{s}}{d T}(T(\bm{x}){-}T)$;
the term $\vec{\tenscomp{n}}^{(1)}({\bm{q}})_{s,s'}{\cdot}\nabla T(\bm{x})$ is, in general, non-diagonal in $s,s'$ and contains the information concerning the deviation of the full solution from the local equilibrium solution --- it is assumed to be of the order of the temperature gradient, as in previous work\cite{fugallo2013ab,cepellotti2016thermal}.
Considering only the first-order term in $\nabla T$, we obtain an equation that is decoupled in its diagonal and off-diagonal parts.
The diagonal part turns out to be the usual LBTE for the populations\cite{fugallo2013ab}:
\begin{equation}
\abovedisplayskip=2mm
\belowdisplayskip=2mm
\begin{split}
\vec{\tenscomp{V}}&(\bm{q})_{s,s} \cdot \left( \frac{\partial \tenscomp{N}^{T}(\bm{q})_{s} }{\partial T } \nabla T(\bm{x})\right)=\\
&=-\frac{1}{\mathcal{V}N_{\rm c}}\sum_{{s''}{\bm{q}''}} \tenscomp{A}^T(\bm{q},{\bm{q}''})_{s,{s''}}\big(\vec{\tenscomp{n}}^{(1)}({{\bm{q}''} })_{{s''},{s''} }{\cdot} \nabla T(\bm{x}) \big);
\end{split}
\raisetag{18mm}
\label{eq:populations_steady_state}
\end{equation}
which can be solved approximately (in the SMA\cite{PhysRevLett.106.045901}) or exactly\cite{fugallo2013ab,cepellotti2016thermal,omini1995iterative,carrete2017almabte,PhysRevLett.110.265506}.
The equation for the coherences ($s{\neq}s'$) is:
\begin{equation}
\abovedisplayskip=2mm
\belowdisplayskip=2mm
\medmuskip=0mu
\thinmuskip=-1mu
\thickmuskip=-1mu
\begin{split}
&\frac{1}{2} \left( \frac{\partial \tenscomp{N}^{T}(\bm{q})_{s} }{\partial T }+\frac{\partial \tenscomp{N}^{T}(\bm{q})_{s'} }{\partial T } \right)\vec{\tenscomp{V}}(\bm{q})_{s,s'} \cdot \nabla T(\bm{x})=\\
&-\left(i\big(\omega(\bm{q})_s-\omega(\bm{q})_{s'}\big)+\frac{\Gamma(\bm{q})_{s}+\Gamma(\bm{q})_{s'}}{2}\right) \vec{\tenscomp{n}}^{(1)}({\bm{q}})_{s,s'}{\cdot}\nabla T(\bm{x}).
\end{split}
\raisetag{17mm}
\label{eq:coherences_steady_state}
\end{equation}
Equation~(\ref{eq:coherences_steady_state}) can be solved exactly and straightforwardly, yielding ($s{\neq} s'$):
\begin{equation}
\abovedisplayskip=1mm
\belowdisplayskip=1mm
\medmuskip=0mu
\thinmuskip=-1mu
\thickmuskip=-1mu
  \begin{split}
&{\tenscomp{n}}^{(1),\alpha}({\bm{q}})_{s,s'}=-\frac{\hbar}{k_{B} {T}^2}{\tenscomp{V}^\alpha}(\bm{q})_{s,s'}\times\\
&\frac{\omega(\bm{q})_{s}\bar{\tenscomp{N}}^{T}\hspace*{-1mm}({\bm{q}})_{s}[\bar{\tenscomp{N}}^{T}\hspace*{-1mm}({\bm{q}})_{s}+1]+\omega(\bm{q})_{s'}\bar{\tenscomp{N}}^{T}\hspace*{-1mm}({\bm{q}})_{s'}[\bar{\tenscomp{N}}^{T}\hspace*{-1mm}({\bm{q}})_{s'}+1]}{2i[\omega(\bm{q})_{s}-\omega(\bm{q})_{s'}]+[\Gamma(\bm{q})_{s}+\Gamma(\bm{q})_{s'}]}.
\end{split}
\label{eq:coherence_solution}
\raisetag{16mm}
\end{equation}

\noindent
We stress that the simplicity of the equation for the coherences~(\ref{eq:coherences_steady_state}) implies that the computational cost for finding its solution is negligible compared to the one for solving exactly the populations' equation~(\ref{eq:populations_steady_state}).\\

\noindent
\textbf{SMA approximation and properties of $\kappa^{\alpha\beta}$}\\
In the kinetic regime\cite{cepellotti2015phonon}, a good estimate of the Peierls' conductivity $\kappa^{\alpha\beta}_{\rm P}$ appearing in equation~(\ref{eq:thermal_conductivity_final_sum}) is given by the SMA approximation, \textit{i.e.} $\kappa^{\alpha\beta}_{\rm P}{\simeq}\kappa^{\alpha\beta}_{_{\rm P,SMA}}$ where
\begin{equation}
\medmuskip=-1mu
\thinmuskip=-2mu
\thickmuskip=0mu
  \kappa^{\alpha\beta}_{_{\rm P,SMA}}\hspace*{-1mm}{=}\hspace*{-0.5mm}\frac{\hbar^2}{k_{B} {T}^2}\hspace*{-0.5mm}\frac{1}{\mathcal{V}N_{\rm c}}\hspace*{-1mm}\sum\limits_{\bm{q},s}\frac{\omega^2\hspace*{-0.5mm}(\bm{q})\hspace*{-0.4mm}_{s}\bar{\tenscomp{N}}^{T}\hspace*{-0.6mm}({\bm{q}})_{\hspace*{-0.5mm}s}[\bar{\tenscomp{N}}^{T}\hspace*{-0.5mm}({\bm{q}})_{\hspace*{-0.5mm}s}\hspace*{-0.5mm}{+}1]}{\Gamma(\bm{q})_{\hspace*{-0.5mm}s}}\hspace*{-0.5mm}{\tenscomp{V}^\alpha\hspace*{-0.5mm}(\hspace*{-0.2mm}\bm{q}\hspace*{-0.2mm})}_{\hspace*{-0.5mm}s\hspace*{-0.4mm},s}\hspace*{-0.7mm}{\tenscomp{V}}^\beta\hspace*{-0.5mm}(\hspace*{-0.2mm}\bm{q}\hspace*{-0.2mm})_{\hspace*{-0.5mm}s\hspace*{-0.4mm},s}.
  \label{eq:SMA_approx}
\end{equation}
Within the SMA approximation, 
the total thermal conductivity $\kappa^{\alpha \beta}$ is approximated by $\kappa^{\alpha \beta}_{_{\rm SMA}}{=}\kappa^{\alpha\beta}_{_{\rm P,SMA}}{+}\kappa^{\alpha\beta}_{_{\rm C}}$, which can be written in the following compact form:
\begin{widetext}
\begin{equation}
\begin{split}
\kappa^{\alpha \beta}_{_{\rm SMA}}=\frac{\hbar^2}{k_{B} {T}^2}\frac{1}{\mathcal{V}N_{\rm c}}\sum_{\bm{q}}\sum_{s,s'}&\frac{\omega(\bm{q})_{s}+\omega(\bm{q})_{s'}}{2}{\tenscomp{V}^\alpha}(\bm{q})_{s,s'}{\tenscomp{V}}^\beta(\bm{q})_{s',s}\times\\
&\times\frac{\omega(\bm{q})_{s}\bar{\tenscomp{N}}^{T}({\bm{q}})_{s}[\bar{\tenscomp{N}}^{T}({\bm{q}})_{s}+1]+\omega(\bm{q})_{s'}\bar{\tenscomp{N}}^{T}({\bm{q}})_{s'}[\bar{\tenscomp{N}}^{T}({\bm{q}})_{s'}+1]}{4[\omega(\bm{q})_{s'}-\omega(\bm{q})_{s}]^2+[\Gamma(\bm{q})_{s}+\Gamma(\bm{q})_{s'}]^2}[\Gamma(\bm{q})_{s}+\Gamma(\bm{q})_{s'}];\label{eq:thermal_conductivity_final_sum_symmetric}
\end{split}
\end{equation}
\end{widetext}
where we stress that the summation over $s,s'$ in equation~(\ref{eq:thermal_conductivity_final_sum_symmetric}) includes the diagonal terms $s{=}s'$ (in contrast, these terms are excluded from the summation in equation~(\ref{eq:thermal_conductivity_final_sum})).
We highlight how the coherences' conductivity $\kappa^{\alpha \beta}_{{\rm C }}$ remains exact (\textit{i.e.} as in equation~(\ref{eq:thermal_conductivity_final_sum})) also in the context of the SMA approximation.
This can be understood from equation~(\ref{eq:scattering_operator}), where it is apparent that the coherences are affected only by scattering events that yield a decrease of the off-diagonal ($s{\neq}s'$) Wigner distribution\cite{cohen2004atom} $\tenscomp{N}(\bm{R},\bm{q},t)_{s,s'}$  (depopulation). Therefore, when the scattering events that yield an increase of the populations are neglected in the SMA approximation, the coherence equation --- affected by depopulation only --- is left unchanged\cite{Simoncelli_full_2018} and thus yields the same value for $\kappa^{\alpha \beta}_{{\rm C }}$ reported in equation~(\ref{eq:thermal_conductivity_final_sum}).

As a final remark for this section, it is worth mentioning that the equations for $\kappa^{xx}_{\rm P}$ in the work of Fugallo \textit{et al.}\cite{fugallo2013ab} are recovered by choosing the eigenstates of $\mathrm{H}^{\rm har}$ in the degenerate subspaces to be eigenstates of $\tens{V}^{x}(\bm{q})$ and neglecting the term $\kappa^{\alpha\beta}_{\rm C}$.\\[10mm]

\noindent
\textbf{The limit of a harmonic crystal with infinite unit cell: Allen-Feldman equation.} 
Here we show that in the limit of a harmonic crystal with an infinite unit cell (\textit{i.e.} in the case of a harmonic glass, whose Brillouin zone is reduced to the $\Gamma$-point), the term $\kappa_{\rm C}^{\alpha\beta}$ mathematically reduces to the Allen-Feldman expression for the thermal conductivity of harmonic glasses\cite{allen1989thermal,allen1993thermal}:
\begin{equation}
  \label{eq:Allen}
  \kappa_A=\frac{1}{V_{\rm t}}C_{i}(T)D_{i}(T);
\end{equation}
where $i{=}1,{\dots}, 3{\cdot} N_{\rm at}^{\rm tot}$ with $N_{\rm at}^{\rm tot}$ the total number of atoms in the (ideally infinite) single unit cell of the glass and $V_{\rm t}$ its (ideally infinite) volume. 
The mode specific heat $C_{i}(T)$ is:\\[-6mm]
\begin{equation}
  C_{i}(T)={\left(\frac{\hbar \omega_{i}^2 }{V_{\rm t}T}\right)
\left(-\frac{\partial \bar{N}_{i} }{\partial \omega_{i}}\right)};
\label{eq:spec_heat}
\end{equation}
where $\bar{N}_{i}{=} \big({\exp\big[\hbar\omega_i/(k_B T) \big]{-}1}\big)^{-1}$ is the Bose-Einstein distribution and the mode diffusivity $D_{i}(T)$ is:
\begin{equation}
  D_{i}(T)=\frac{\pi V_{\rm t}^2}{3\hbar^2 \omega_{i}^2 }\frac{\hbar^2}{V^2}\sum_{j\neq i}\left|\vec{\tenscomp{V}}_{\text{\tiny{AA}}}(\bm{q})_{i,j}\frac{\omega_{i}+\omega_{j}}{2}\right|^2\delta(\omega_{i}-\omega_{j}).
  \label{eq:diffusivity}
\end{equation}
$\tenscomp{V}^\beta_{\text{\tiny{AA}}}(\bm{q})_{s,s'}$ is the velocity operator defined by Auerbach and Allen\cite{PhysRevB.29.2884}, which is related to the velocity operator~(\ref{eq:vel_op}) derived in this work:
\begin{equation}
  \tenscomp{V}^\beta_{\text{\tiny{AA}}}(\bm{q})_{s,s'} =\frac{\omega(\bm{q})_s+\omega(\bm{q})_{s'}}{{2 \sqrt{\omega(\bm{q})_s\omega(\bm{q})_{s'}}}} \tenscomp{V}^\beta(\bm{q})_{s,s'}.
  \label{eq:vel_op_rel}
\end{equation}
The Dirac delta in equation~(\ref{eq:diffusivity}) implies that these two velocity operators in equation~(\ref{eq:vel_op_rel}) are equivalent in the computation of the Allen-Feldman diffusivities~(\ref{eq:diffusivity}).\\
Equation~(\ref{eq:Allen}) is derived under the hypothesis of extreme disorder --- \textit{i.e.} the phonon wave-packets do not to propagate far enough to sample the periodicity of the medium, rendering it impossible to assign them a wave vector or a group velocity.
In practice, this  
corresponds to neglecting the thermal conductivity contributions coming from the diagonal elements of the velocity operator (otherwise, at the harmonic order, these would yield a divergent thermal conductivity\cite{feldman1993thermal}). 
Within such extreme disorder assumption, we are left with the coherences' term and we will show now that $\frac{1}{3}\kappa_C^{\alpha\alpha}$ is equivalent to the Allen-Feldman formula~(\ref{eq:Allen}).
Let us start by writing explicitly part of equation~(\ref{eq:spec_heat}):
\begin{equation}
-\frac{\partial \bar{N}(\bm{q})_s }{\partial \omega(\bm{q})_{s}} \frac{\hbar \omega({\bm{q})_s} }{V_{\rm t} T}=
\frac{1}{\mathcal{V}N_{\rm c}}\frac{\hbar^2  }{k_B T^2}\omega(\bm{q})_{s}\bar{N}(\bm{q})_s(\bar{N}(\bm{q})_s+1);
\end{equation}
where we have used $V_{\rm t}{=}\mathcal{V}N_{\rm c}$. Taking first the limit $\Gamma(\bm{q})_{s}{\to} \eta\;\forall\hspace*{1mm} (s,\;\bm{q})$ and then $\eta{\to} 0$, we have that the coherences' term in equation~(\ref{eq:thermal_conductivity_final_sum}) becomes:
\begin{equation}
\begin{split}
\kappa^{\alpha \beta}_{\rm dis}{=}&
 {-}\sum_{\bm{q},s\neq s'}
\frac{\hbar \omega(\bm{q})_{s}^2 }{V_{\rm t}T}
\frac{\partial \bar{N}(\bm{q})_{s} }{\partial \omega(\bm{q})_{s}}\frac{\pi}{\omega(\bm{q})_{s}} \delta(\omega(\bm{q})_{s'}{-}\omega(\bm{q})_{s})\\
 &{\times}\frac{\omega(\bm{q})_{s}+\omega(\bm{q})_{s'}}{2}{\tenscomp{V}^\alpha}(\bm{q})_{s,s'}{\tenscomp{V}}^\beta(\bm{q})_{s',s};
\end{split}
\label{eq:kappa_dis}
\raisetag{6mm}
\end{equation}
where we have obtained the delta function as the limit of a Lorentzian for ${\Gamma( \bm{q})_{s}{\to} \eta {\to} 0}$. 
It follows that, defining $\kappa_A{=} {\kappa^{\alpha\alpha}_{\rm dis}}/{3}$ and in the case of a Brillouin zone reduced to the $\Gamma$-point, we recover the Allen-Feldman formula:
\begin{equation}
\kappa_A{=}\frac{1}{3}\sum_{s'\neq s}
-\frac{\hbar \omega_{s}^2 }{V_{\rm t}T}\frac{\partial \bar{N}_{s} }{\partial \omega_{s}}\frac{\pi}{\omega^2_{s}} \delta(\omega_{s'}-\omega_{s})
 \left(\hspace*{-1mm}\frac{\omega_{s}+\omega_{s'}}{2}\hspace*{-1mm}\right)^{\hspace*{-1mm} 2}{\tenscomp{V}}_{s,s'}^\alpha{\tenscomp{V}}^\alpha_{s',s};
\label{eq:AF_formula}
\end{equation}
where $\omega_{s}{=}\omega(\bm{q}{=}\bm{0})_{s}$, $\bar{N}_{s}{=}\bar{N}(\bm{q}{=}\bm{0})_{s}$ and $\vec{\tens{V}}{=}\vec{\tens{V}}(\bm{q}{=}\bm{0})$.\\[2mm] 

\begin{figure}[t]
\vspace*{-2mm}
  \centering
  \includegraphics[width=\columnwidth]{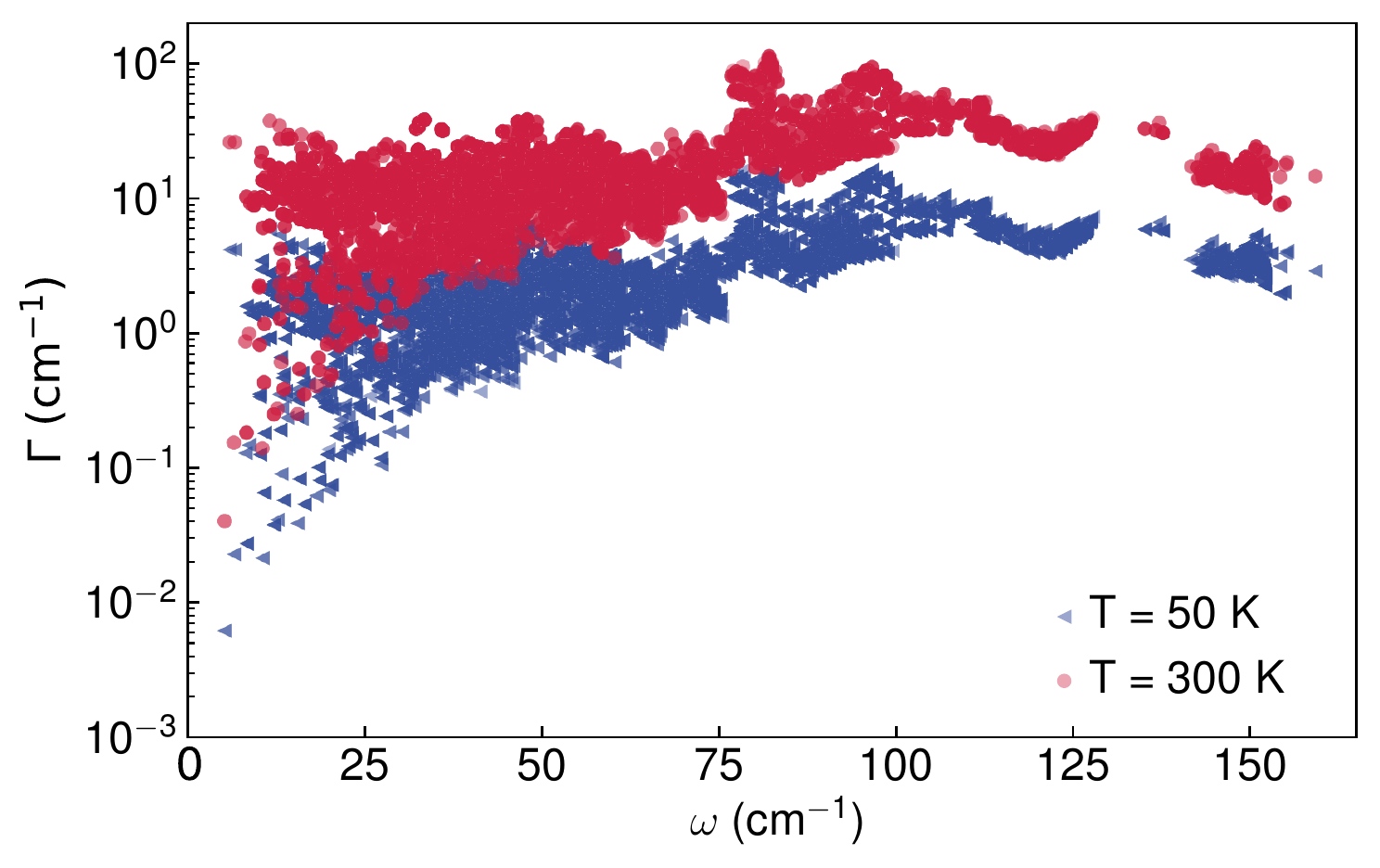}\\[-4mm]
  \caption{\textbf{Phonon linewidths distribution of CsPbBr$_3$}. The linewidths at 50~K and 300~K are respectively reported in blue  and red. An increase in temperature results in a shift of the linewidths towards larger values.}
  \label{fig:ph_Lw}
\end{figure}

\noindent
\textbf{Phonon linewidths distribution.} 
Figure~(\ref{fig:ph_Lw}) shows the distribution of the phonon linewidths for CsPbBr$_3$ at different temperatures. 
The distribution spans a range of about 4 orders of magnitude (from about $10^{-2}$ cm$^{-1}$ to about $10^{2}$ cm$^{-1}$); consequently, couplings between vibrational eigenstates having very different frequencies are allowed, as shown in figure~(\ref{fig:cond_mech}\textbf{d,e}). 
Increasing the temperature yields an increase of the phonon linewidths, with consequent higher $\kappa^{\alpha\beta}_{\rm C}$ --- since more couplings are allowed --- and lower $\kappa^{\alpha\beta}_{\rm B}$.\\[10mm]

\noindent
\textbf{Computational details.}
The experimental structure of orthorhombic CsPbBr$_3$\cite{exp_cell_param} with \textit{Pnma} space group (Crystallographic Open Database\cite{COD_database} id: 4510745) is first converted into an equivalent structure with \textit{Pbnm} space group using VESTA\cite{momma2011vesta}, to allow a direct comparison with other references on this material\cite{wang2018cation,Miyatae1701217}.
The first-principles equilibrium lattice parameters are then determined using Quantum ESPRESSO\cite{giannozzi2017advanced}, the PBEsol exchange-correlation functional and GBRV\cite{garrity2014pseudopotentials} pseudopotentials. 
The PBEsol functional is selected on the basis of its good agreement between  first-principles and experimental lattice parameters\cite{lee2017ultralow,exp_cell_param} (see table~(\ref{tab:cell_param})) and its capability to accurately describe the vibrational properties of inorganic halide perovskites\cite{lee2017ultralow}.
Kinetic energy cutoffs of 50 and 400 Ry are used for the wave functions and the charge density. The Brillouin zone is integrated with a Monkhorst-Pack mesh of $5{\times}5{\times}4$ points, with a (1,1,1) shift. 
Second-order force constants are computed on a $4{\times}4{\times}4$ mesh using density-functional perturbation theory\cite{RevModPhys.73.515} as implemented in Quantum ESPRESSO.
In order to obtain the third-order force constants using the PBEsol functional, the finite-difference method implemented in ShengBTE\cite{li2014shengbte} is used, together with the interconversion software from ShengBTE to Quantum ESPRESSO available in the D3Q package\cite{paulatto2015first}.
Third-order force constants are computed on a $2{\times}2{\times}2$ mesh and considering interactions up to the third-nearest neighbors.
All the force constants are Fourier interpolated on  $8{\times}7{\times}5$ grid for the thermal conductivity calculations. In the latter, natural-abundance isotopic scattering\cite{PhysRevLett.106.045901,tamura1983isotope} 
is considered in addition to third-order anharmonicity\cite{carrete2017almabte,fugallo2013ab,phono3py,paulatto2013anharmonic,PhonTS,alamode}. 
From the force constants, the phonon frequencies and the full non-diagonal anharmonic scattering operator are computed as in previous work\cite{fugallo2013ab}. 
The anharmonic scattering operator is then modified to account for the coherences' term reported in equation~(\ref{eq:scattering_operator}).
In figure~(\ref{fig:k_vs_T}), a smearing $\sigma$ of $2\;{\rm cm^{-1}}$ is used to compute the delta function appearing in the scattering operator. Results are converged with respect to the choice of this parameter: changing $\sigma$ to $1\;{\rm cm^{-1}}$ and $5\;{\rm cm^{-1}}$ produces populations, coherences and total thermal conductivities with a maximum relative deviation of  $2.6\%$ from the data of figure~(\ref{fig:k_vs_T}). 
These latter refer to the single-mode relaxation time approximation (SMA); the exact thermal conductivity is always compatible within $0.7 \%$ to its SMA value.
Performing the thermal conductivity calculations using 
force constants that are Fourier interpolated on a $4{\times}4{\times}3$ mesh
produces data compatible with the ones reported in figure~(\ref{fig:k_vs_T}), which use a $8{\times}7{\times}5$ mesh.
In particular, the maximum relative deviation are respectively 1.8$\%$ for the populations in the SMA and 0.9$\%$ for the coherences' thermal conductivity.
As anticipated in the main text, the data shown in figure~(\ref{fig:k_vs_T}) for $\kappa^{\alpha\beta}_{\rm P}$ are compatible with the theoretical results presented in  Ref.\cite{lee2017ultralow} (and related Supplementary Material). In particular, the direction $xx$ in this work corresponds to the $[010]$ direction in Ref.\cite{lee2017ultralow}. Experimental data from Ref.\cite{lee2017ultralow} do not allow to evaluate the effect of boundary scattering, as they refer to nanowires having roughly the same size, and are smaller than all the nanowires used in Ref.\cite{wang2018cation}. Therefore, we compared the bulk thermal conductivity predictions of equation~(\ref{eq:thermal_conductivity_final_sum}) with the experimental data from Ref.\cite{wang2018cation}.
In the calculation of the data reported in figure~(\ref{fig:k_vs_T}), temperature effects are accounted through the Bose-Einstein distributions entering in the scattering operator. Accounting for the  renormalization of the force constants due to temperature goes beyond the conceptual scope of this work but it has been addressed recently by several studies\cite{PhysRevB.96.014111,PhysRevLett.120.105901,PhysRevB.94.020303,PhysRevB.98.024301,PhysRevLett.112.058501,PhysRevB.98.085205,PhysRevB.91.214310,aseginolaza2018phonon,PhysRevB.84.180301,PhysRevB.88.144301,PhysRevB.87.104111,PhysRevLett.100.095901}.
Calculations for diamond and silicon are performed using a $26{\times}26{\times}26$ q-point grid and a smearing of $8\;{\rm cm}^{-1}$ and $4\;{\rm cm}^{-1}$ respectively. Details on the first-principles calculation of the force constants for these materials can be found in Refs.\cite{fugallo2013ab,cepellotti2016thermal}.
In figure~(\mbox{\ref{fig:cond_mech}\textbf{{d,e}}}), contributions to $\kappa^{xx}_{\rm{C,DOS}(2)}$ smaller than $4{\cdot}10^{-6}\;\mathrm{\frac{W{\cdot}cm^2}{m\cdot K}}$ are not reported for graphical clarity.
\begin{table}[h]
  \caption{Lattice parameters of CsPbBr$_3$ in the orthorhombic phase, obtained by first-principles simulations or experiments\cite{lee2017ultralow,Miyatae1701217}. Inset: unit cell of orthorhombic CsPbBr$_3$ visualized with VESTA\cite{momma2011vesta}; the directions $\hat{x},\hat{y}$ and $\hat{z}$ discussed e.g. in figure~(\ref{fig:k_vs_T}) refer respectively to lattice vectors $\bm{a}_1$, $\bm{a}_2$ and $\bm{a}_3$.}
  \label{tab:cell_param}
  \centering
  \begin{tabular}{l|c|c|c|c}
  \hline

  \hline
  & ${a}_1$ [\AA] & ${a}_2$ [\AA] & ${a}_3$ [\AA] &\multirow{4}*{\includegraphics[width=1.9cm]{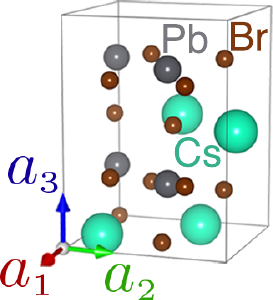}}\\
  \cline{1-4}
    Present (th) & 7.963 & 8.389 & 11.632 & \\
   \cline{1-4}
    Ref.\cite{lee2017ultralow} (th) & 7.990 & 8.416 & 11.674 & \\
   \cline{1-4}
    Ref.\cite{exp_cell_param} (exp) & 8.198  & 8.244  & 11.735  &   \\
   \cline{1-4}
    Ref.\cite{Miyatae1701217} (exp) & 8.223 & 8.243 & 11.761 &   \\
  \hline

  \hline
  \end{tabular}
\end{table}

\noindent
{\sf{\textbf{{Data availability.}}}}  Raw data were generated using the SCITAS High Performance Computing facility at the {\'E}cole Polytechnique F{\'e}d{\'e}rale de Lausanne. Derived data supporting the findings of this study are available at \url{https://archive.materialscloud.org}.\\[1mm]

\noindent
{\sf{\textbf{{Code availability.}}}} Quantum ESPRESSO is available at \url{www.quantum-espresso.org}; the scripts related to the computation of the third order force constants using the finite-difference method are available at \url{bitbucket.org/sousaw/thirdorder}; the D3Q package for Quantum ESPRESSO is available at \url{sourceforge.net/projects/d3q/}. The custom code developed in this work will be made available in a next release of the D3Q package. 

\bibliographystyle{naturemag}
\bibliography{biblio.bbl}

\begin{thebibliography}{10}
\expandafter\ifx\csname url\endcsname\relax
  \def\url#1{\texttt{#1}}\fi
\expandafter\ifx\csname urlprefix\endcsname\relax\def\urlprefix{URL }\fi
\providecommand{\bibinfo}[2]{#2}
\providecommand{\eprint}[2][]{\url{#2}}

\bibitem{peierls1929kinetischen}
\bibinfo{author}{Peierls, R.}
\newblock \bibinfo{title}{{ Zur Kinetischen Theorie der W{\"a}rmeleitung in
  Kristallen}}.
\newblock \emph{\bibinfo{journal}{Ann. Phys. (N.Y.)}}
  \textbf{\bibinfo{volume}{395}}, \bibinfo{pages}{1055--1101}
  (\bibinfo{year}{1929}).

\bibitem{allen1989thermal}
\bibinfo{author}{Allen, P.~B.} \& \bibinfo{author}{Feldman, J.~L.}
\newblock \bibinfo{title}{{Thermal Conductivity of Glasses: Theory and
  Application to Amorphous Si}}.
\newblock \emph{\bibinfo{journal}{Phys. Rev. Lett.}}
  \textbf{\bibinfo{volume}{62}}, \bibinfo{pages}{645--648}
  (\bibinfo{year}{1989}).

\bibitem{sun2010lattice}
\bibinfo{author}{Sun, T.} \& \bibinfo{author}{Allen, P.~B.}
\newblock \bibinfo{title}{{Lattice thermal conductivity: Computations and
  theory of the high-temperature breakdown of the phonon-gas model}}.
\newblock \emph{\bibinfo{journal}{Phys. Rev. B}} \textbf{\bibinfo{volume}{82}},
  \bibinfo{pages}{224305} (\bibinfo{year}{2010}).

\bibitem{li2015ultralow}
\bibinfo{author}{Li, W.} \& \bibinfo{author}{Mingo, N.}
\newblock \bibinfo{title}{{Ultralow lattice thermal conductivity of the fully
  filled skutterudite YbFe$_{4}$Sb$_{12}$ due to the flat avoided-crossing
  filler modes}}.
\newblock \emph{\bibinfo{journal}{Phys. Rev. B}} \textbf{\bibinfo{volume}{91}},
  \bibinfo{pages}{144304} (\bibinfo{year}{2015}).

\bibitem{wang2018cation}
\bibinfo{author}{Wang, Y.} \emph{et~al.}
\newblock \bibinfo{title}{{Cation Dynamics Governed Thermal Properties of Lead
  Halide Perovskite Nanowires}}.
\newblock \emph{\bibinfo{journal}{Nano Lett.}} \textbf{\bibinfo{volume}{18}},
  \bibinfo{pages}{2772--2779} (\bibinfo{year}{2018}).

\bibitem{lee2017ultralow}
\bibinfo{author}{Lee, W.} \emph{et~al.}
\newblock \bibinfo{title}{{Ultralow thermal conductivity in all-inorganic
  halide perovskites}}.
\newblock \emph{\bibinfo{journal}{Proc. Natl. Acad. Sci. U. S. A.}}
  \textbf{\bibinfo{volume}{114}}, \bibinfo{pages}{8693--8697}
  (\bibinfo{year}{2017}).

\bibitem{chen2015twisting}
\bibinfo{author}{Chen, X.} \emph{et~al.}
\newblock \bibinfo{title}{{Twisting phonons in complex crystals with
  quasi-one-dimensional substructures}}.
\newblock \emph{\bibinfo{journal}{Nat. Commun.}} \textbf{\bibinfo{volume}{6}},
  \bibinfo{pages}{6723} (\bibinfo{year}{2015}).

\bibitem{voneshen2013suppression}
\bibinfo{author}{Voneshen, D.} \emph{et~al.}
\newblock \bibinfo{title}{{Suppression of thermal conductivity by rattling
  modes in thermoelectric sodium cobaltate}}.
\newblock \emph{\bibinfo{journal}{Nat. Mater.}} \textbf{\bibinfo{volume}{12}},
  \bibinfo{pages}{1028--1032} (\bibinfo{year}{2013}).

\bibitem{lory2017direct}
\bibinfo{author}{Lory, P.-F.} \emph{et~al.}
\newblock \bibinfo{title}{{Direct measurement of individual phonon lifetimes in
  the clathrate compound Ba 7.81 Ge 40.67 Au 5.33}}.
\newblock \emph{\bibinfo{journal}{Nat. Commun.}} \textbf{\bibinfo{volume}{8}},
  \bibinfo{pages}{491} (\bibinfo{year}{2017}).

\bibitem{PhysRevB.96.214202}
\bibinfo{author}{Weathers, A.} \emph{et~al.}
\newblock \bibinfo{title}{{Glass-like thermal conductivity in nanostructures of
  a complex anisotropic crystal}}.
\newblock \emph{\bibinfo{journal}{Phys. Rev. B}} \textbf{\bibinfo{volume}{96}},
  \bibinfo{pages}{214202} (\bibinfo{year}{2017}).

\bibitem{PhysRevLett.106.045901}
\bibinfo{author}{Garg, J.}, \bibinfo{author}{Bonini, N.},
  \bibinfo{author}{Kozinsky, B.} \& \bibinfo{author}{Marzari, N.}
\newblock \bibinfo{title}{{Role of Disorder and Anharmonicity in the Thermal
  Conductivity of Silicon-Germanium Alloys: A First-Principles Study}}.
\newblock \emph{\bibinfo{journal}{Phys. Rev. Lett.}}
  \textbf{\bibinfo{volume}{106}}, \bibinfo{pages}{045901}
  (\bibinfo{year}{2011}).

\bibitem{omini1995iterative}
\bibinfo{author}{Omini, M.} \& \bibinfo{author}{Sparavigna, A.}
\newblock \bibinfo{title}{{An iterative approach to the phonon Boltzmann
  equation in the theory of thermal conductivity}}.
\newblock \emph{\bibinfo{journal}{Physica B: Condens. Matter}}
  \textbf{\bibinfo{volume}{212}}, \bibinfo{pages}{101 -- 112}
  (\bibinfo{year}{1995}).

\bibitem{carrete2017almabte}
\bibinfo{author}{Carrete, J.} \emph{et~al.}
\newblock \bibinfo{title}{{almaBTE : A solver of the space–time dependent
  Boltzmann transport equation for phonons in structured materials}}.
\newblock \emph{\bibinfo{journal}{Comput. Phys. Commun.}}
  \textbf{\bibinfo{volume}{220}}, \bibinfo{pages}{351 -- 362}
  (\bibinfo{year}{2017}).

\bibitem{fugallo2013ab}
\bibinfo{author}{Fugallo, G.}, \bibinfo{author}{Lazzeri, M.},
  \bibinfo{author}{Paulatto, L.} \& \bibinfo{author}{Mauri, F.}
\newblock \bibinfo{title}{{Ab initio variational approach for evaluating
  lattice thermal conductivity}}.
\newblock \emph{\bibinfo{journal}{Phys. Rev. B}} \textbf{\bibinfo{volume}{88}},
  \bibinfo{pages}{045430} (\bibinfo{year}{2013}).

\bibitem{PhysRevLett.110.265506}
\bibinfo{author}{Chaput, L.}
\newblock \bibinfo{title}{{Direct Solution to the Linearized Phonon Boltzmann
  Equation}}.
\newblock \emph{\bibinfo{journal}{Phys. Rev. Lett.}}
  \textbf{\bibinfo{volume}{110}}, \bibinfo{pages}{265506}
  (\bibinfo{year}{2013}).

\bibitem{cepellotti2016thermal}
\bibinfo{author}{Cepellotti, A.} \& \bibinfo{author}{Marzari, N.}
\newblock \bibinfo{title}{{Thermal Transport in Crystals as a Kinetic Theory of
  Relaxons}}.
\newblock \emph{\bibinfo{journal}{Phys. Rev. X}} \textbf{\bibinfo{volume}{6}},
  \bibinfo{pages}{041013} (\bibinfo{year}{2016}).

\bibitem{cepellotti2015phonon}
\bibinfo{author}{Cepellotti, A.} \emph{et~al.}
\newblock \bibinfo{title}{{{Phonon hydrodynamics in two-dimensional
  materials}}}.
\newblock \emph{\bibinfo{journal}{Nat. Commun.}} \textbf{\bibinfo{volume}{6}},
  \bibinfo{pages}{6400} (\bibinfo{year}{2015}).

\bibitem{hardy1963energy}
\bibinfo{author}{Hardy, R.~J.}
\newblock \bibinfo{title}{{Energy-Flux Operator for a Lattice}}.
\newblock \emph{\bibinfo{journal}{Phys. Rev.}} \textbf{\bibinfo{volume}{132}},
  \bibinfo{pages}{168--177} (\bibinfo{year}{1963}).

\bibitem{allen1999diffusons}
\bibinfo{author}{Allen, P.~B.}, \bibinfo{author}{Feldman, J.~L.},
  \bibinfo{author}{Fabian, J.} \& \bibinfo{author}{Wooten, F.}
\newblock \bibinfo{title}{{Diffusons, locons and propagons: Character of atomie
  yibrations in amorphous Si}}.
\newblock \emph{\bibinfo{journal}{Philos. Mag. B}}
  \textbf{\bibinfo{volume}{79}}, \bibinfo{pages}{1715--1731}
  (\bibinfo{year}{1999}).

\bibitem{lv2016non}
\bibinfo{author}{Lv, W.} \& \bibinfo{author}{Henry, A.}
\newblock \bibinfo{title}{{Non-negligible contributions to thermal conductivity
  from localized modes in amorphous silicon dioxide}}.
\newblock \emph{\bibinfo{journal}{Scientific reports}}
  \textbf{\bibinfo{volume}{6}}, \bibinfo{pages}{35720} (\bibinfo{year}{2016}).

\bibitem{ziman1960electrons}
\bibinfo{author}{Ziman, J.~M.}
\newblock \emph{\bibinfo{title}{{Electrons and phonons: the theory of transport
  phenomena in solids}}} (\bibinfo{publisher}{Oxford university press},
  \bibinfo{year}{1960}).

\bibitem{tamura1983isotope}
\bibinfo{author}{Tamura, S.-i.}
\newblock \bibinfo{title}{{Isotope scattering of dispersive phonons in Ge}}.
\newblock \emph{\bibinfo{journal}{Phys. Rev. B}} \textbf{\bibinfo{volume}{27}},
  \bibinfo{pages}{858--866} (\bibinfo{year}{1983}).

\bibitem{gebauer2004kinetic}
\bibinfo{author}{Gebauer, R.} \& \bibinfo{author}{Car, R.}
\newblock \bibinfo{title}{{Kinetic theory of quantum transport at the
  nanoscale}}.
\newblock \emph{\bibinfo{journal}{Phys. Rev. B}} \textbf{\bibinfo{volume}{70}},
  \bibinfo{pages}{125324} (\bibinfo{year}{2004}).

\bibitem{frensley1990boundary}
\bibinfo{author}{Frensley, W.~R.}
\newblock \bibinfo{title}{{Boundary conditions for open quantum systems driven
  far from equilibrium}}.
\newblock \emph{\bibinfo{journal}{Rev. Mod. Phys.}}
  \textbf{\bibinfo{volume}{62}}, \bibinfo{pages}{745--791}
  (\bibinfo{year}{1990}).

\bibitem{cohen2004atom}
\bibinfo{author}{Cohen-Tannoudji, C.}, \bibinfo{author}{Dupont-Roc, J.},
  \bibinfo{author}{Grynberg, G.} \& \bibinfo{author}{Thickstun, P.}
\newblock \emph{\bibinfo{title}{{Atom-photon interactions: basic processes and
  applications}}} (\bibinfo{publisher}{Wiley Online Library},
  \bibinfo{year}{2004}).

\bibitem{PhysRevB.46.6131}
\bibinfo{author}{Cahill, D.~G.}, \bibinfo{author}{Watson, S.~K.} \&
  \bibinfo{author}{Pohl, R.~O.}
\newblock \bibinfo{title}{{Lower limit to the thermal conductivity of
  disordered crystals}}.
\newblock \emph{\bibinfo{journal}{Phys. Rev. B}} \textbf{\bibinfo{volume}{46}},
  \bibinfo{pages}{6131--6140} (\bibinfo{year}{1992}).

\bibitem{agne2018minimum}
\bibinfo{author}{Agne, M.~T.}, \bibinfo{author}{Hanus, R.} \&
  \bibinfo{author}{Snyder, G.~J.}
\newblock \bibinfo{title}{{Minimum thermal conductivity in the context of
  diffuson-mediated thermal transport}}.
\newblock \emph{\bibinfo{journal}{Energy Environ. Sci.}}
  \textbf{\bibinfo{volume}{11}}, \bibinfo{pages}{609--616}
  (\bibinfo{year}{2018}).

\bibitem{Clarke2003}
\bibinfo{author}{Clarke, D.~R.}
\newblock \bibinfo{title}{{Materials selection guidelines for low thermal
  conductivity thermal barrier coatings}}.
\newblock \emph{\bibinfo{journal}{Surface and Coatings Technology}}
  \textbf{\bibinfo{volume}{163-164}}, \bibinfo{pages}{67--74}
  (\bibinfo{year}{2003}).

\bibitem{marcolongo2016microscopic}
\bibinfo{author}{Marcolongo, A.}, \bibinfo{author}{Umari, P.} \&
  \bibinfo{author}{Baroni, S.}
\newblock \bibinfo{title}{{Microscopic theory and quantum simulation of atomic
  heat transport}}.
\newblock \emph{\bibinfo{journal}{Nat. Phys.}} \textbf{\bibinfo{volume}{12}},
  \bibinfo{pages}{80} (\bibinfo{year}{2016}).

\bibitem{seyf2017rethinking}
\bibinfo{author}{Seyf, H.~R.} \emph{et~al.}
\newblock \bibinfo{title}{{Rethinking phonons: The issue of disorder}}.
\newblock \emph{\bibinfo{journal}{npj Comp. Mat.}}
  \textbf{\bibinfo{volume}{3}}, \bibinfo{pages}{49} (\bibinfo{year}{2017}).

\bibitem{carbogno2017ab}
\bibinfo{author}{Carbogno, C.}, \bibinfo{author}{Ramprasad, R.} \&
  \bibinfo{author}{Scheffler, M.}
\newblock \bibinfo{title}{{Ab Initio Green-Kubo Approach for the Thermal
  Conductivity of Solids}}.
\newblock \emph{\bibinfo{journal}{Phys. Rev. Lett.}}
  \textbf{\bibinfo{volume}{118}}, \bibinfo{pages}{175901}
  (\bibinfo{year}{2017}).

\bibitem{puligheddu2017first}
\bibinfo{author}{Puligheddu, M.}, \bibinfo{author}{Gygi, F.} \&
  \bibinfo{author}{Galli, G.}
\newblock \bibinfo{title}{{First-principles simulations of heat transport}}.
\newblock \emph{\bibinfo{journal}{Phys. Rev. Materials}}
  \textbf{\bibinfo{volume}{1}}, \bibinfo{pages}{060802} (\bibinfo{year}{2017}).

\bibitem{PhysRevB.86.155433}
\bibinfo{author}{Kan\'e, G.}, \bibinfo{author}{Lazzeri, M.} \&
  \bibinfo{author}{Mauri, F.}
\newblock \bibinfo{title}{{Zener tunneling in the electrical transport of
  quasimetallic carbon nanotubes}}.
\newblock \emph{\bibinfo{journal}{Phys. Rev. B}} \textbf{\bibinfo{volume}{86}},
  \bibinfo{pages}{155433} (\bibinfo{year}{2012}).

\bibitem{Simoncelli_full_2018}
\bibinfo{author}{Simoncelli, M.}, \bibinfo{author}{Marzari, N.} \&
  \bibinfo{author}{Mauri, F.}
\newblock \emph{\bibinfo{journal}{In preparation}} .

\bibitem{RevModPhys.84.1419}
\bibinfo{author}{Marzari, N.}, \bibinfo{author}{Mostofi, A.~A.},
  \bibinfo{author}{Yates, J.~R.}, \bibinfo{author}{Souza, I.} \&
  \bibinfo{author}{Vanderbilt, D.}
\newblock \bibinfo{title}{{Maximally localized Wannier functions: Theory and
  applications}}.
\newblock \emph{\bibinfo{journal}{Rev. Mod. Phys.}}
  \textbf{\bibinfo{volume}{84}}, \bibinfo{pages}{1419--1475}
  (\bibinfo{year}{2012}).

\bibitem{hardy1970phonon}
\bibinfo{author}{Hardy, R.~J.}
\newblock \bibinfo{title}{{Phonon Boltzmann equation and second sound in
  solids}}.
\newblock \emph{\bibinfo{journal}{Phys. Rev. B}} \textbf{\bibinfo{volume}{2}},
  \bibinfo{pages}{1193} (\bibinfo{year}{1970}).

\bibitem{allen1993thermal}
\bibinfo{author}{Allen, P.~B.} \& \bibinfo{author}{Feldman, J.~L.}
\newblock \bibinfo{title}{{Thermal conductivity of disordered harmonic
  solids}}.
\newblock \emph{\bibinfo{journal}{Phys. Rev. B}} \textbf{\bibinfo{volume}{48}},
  \bibinfo{pages}{12581--12588} (\bibinfo{year}{1993}).

\bibitem{PhysRevB.29.2884}
\bibinfo{author}{Auerbach, A.} \& \bibinfo{author}{Allen, P.~B.}
\newblock \bibinfo{title}{{Universal high-temperature saturation in phonon and
  electron transport}}.
\newblock \emph{\bibinfo{journal}{Phys. Rev. B}} \textbf{\bibinfo{volume}{29}},
  \bibinfo{pages}{2884--2890} (\bibinfo{year}{1984}).

\bibitem{feldman1993thermal}
\bibinfo{author}{Feldman, J.~L.}, \bibinfo{author}{Kluge, M.~D.},
  \bibinfo{author}{Allen, P.~B.} \& \bibinfo{author}{Wooten, F.}
\newblock \bibinfo{title}{{Thermal conductivity and localization in glasses:
  Numerical study of a model of amorphous silicon}}.
\newblock \emph{\bibinfo{journal}{Phys. Rev. B}} \textbf{\bibinfo{volume}{48}},
  \bibinfo{pages}{12589--12602} (\bibinfo{year}{1993}).

\bibitem{exp_cell_param}
\bibinfo{author}{{Stoumpos, {Constantinos C.} and Malliakas, {Christos D.} and
  Peters, {John A.} and Zhifu Liu and Maria Sebastian and Jino Im and Chasapis,
  {Thomas C.} and Wibowo, {Arief C.} and Chung, {Duck Young} and Freeman,
  {Arthur J} and Wessels, {Bruce W.} and Kanatzidis, {Mercouri G}}}.
\newblock \bibinfo{title}{{Crystal growth of the perovskite semiconductor
  CsPbBr3: A new material for high-energy radiation detection}}.
\newblock \emph{\bibinfo{journal}{Cryst. Growth Des.}}
  \textbf{\bibinfo{volume}{13}}, \bibinfo{pages}{2722--2727}
  (\bibinfo{year}{2013}).

\bibitem{COD_database}
\bibinfo{author}{Gra{\v{z}}ulis, S.} \emph{et~al.}
\newblock \bibinfo{title}{{Crystallography Open Database {--} an open-access
  collection of crystal structures}}.
\newblock \emph{\bibinfo{journal}{J. Appl. Crystallogr.}}
  \textbf{\bibinfo{volume}{42}}, \bibinfo{pages}{726--729}
  (\bibinfo{year}{2009}).

\bibitem{momma2011vesta}
\bibinfo{author}{Momma, K.} \& \bibinfo{author}{Izumi, F.}
\newblock \bibinfo{title}{{VESTA 3 for three-dimensional visualization of
  crystal, volumetric and morphology data}}.
\newblock \emph{\bibinfo{journal}{J. Appl. Crystallogr.}}
  \textbf{\bibinfo{volume}{44}}, \bibinfo{pages}{1272--1276}
  (\bibinfo{year}{2011}).

\bibitem{Miyatae1701217}
\bibinfo{author}{Miyata, K.} \emph{et~al.}
\newblock \bibinfo{title}{{Large polarons in lead halide perovskites}}.
\newblock \emph{\bibinfo{journal}{Sci. Adv.}} \textbf{\bibinfo{volume}{3}}
  (\bibinfo{year}{2017}).

\bibitem{giannozzi2017advanced}
\bibinfo{author}{Giannozzi, P.} \emph{et~al.}
\newblock \bibinfo{title}{{Advanced capabilities for materials modelling with
  Quantum ESPRESSO}}.
\newblock \emph{\bibinfo{journal}{J. Phys. Condens. Matter}}
  \textbf{\bibinfo{volume}{29}}, \bibinfo{pages}{465901}
  (\bibinfo{year}{2017}).

\bibitem{garrity2014pseudopotentials}
\bibinfo{author}{Garrity, K.~F.}, \bibinfo{author}{Bennett, J.~W.},
  \bibinfo{author}{Rabe, K.~M.} \& \bibinfo{author}{Vanderbilt, D.}
\newblock \bibinfo{title}{{Pseudopotentials for high-throughput DFT
  calculations}}.
\newblock \emph{\bibinfo{journal}{Comput. Mater. Sci.}}
  \textbf{\bibinfo{volume}{81}}, \bibinfo{pages}{446 -- 452}
  (\bibinfo{year}{2014}).

\bibitem{RevModPhys.73.515}
\bibinfo{author}{Baroni, S.}, \bibinfo{author}{de~Gironcoli, S.},
  \bibinfo{author}{Dal~Corso, A.} \& \bibinfo{author}{Giannozzi, P.}
\newblock \bibinfo{title}{{Phonons and related crystal properties from
  density-functional perturbation theory}}.
\newblock \emph{\bibinfo{journal}{Rev. Mod. Phys.}}
  \textbf{\bibinfo{volume}{73}}, \bibinfo{pages}{515--562}
  (\bibinfo{year}{2001}).

\bibitem{li2014shengbte}
\bibinfo{author}{Li, W.}, \bibinfo{author}{Carrete, J.},
  \bibinfo{author}{Katcho, N.~A.} \& \bibinfo{author}{Mingo, N.}
\newblock \bibinfo{title}{{ShengBTE: A solver of the Boltzmann transport
  equation for phonons}}.
\newblock \emph{\bibinfo{journal}{Comput. Phys. Commun.}}
  \textbf{\bibinfo{volume}{185}}, \bibinfo{pages}{1747 -- 1758}
  (\bibinfo{year}{2014}).

\bibitem{paulatto2015first}
\bibinfo{author}{Paulatto, L.}, \bibinfo{author}{Errea, I.},
  \bibinfo{author}{Calandra, M.} \& \bibinfo{author}{Mauri, F.}
\newblock \bibinfo{title}{{First-principles calculations of phonon frequencies,
  lifetimes, and spectral functions from weak to strong anharmonicity: The
  example of palladium hydrides}}.
\newblock \emph{\bibinfo{journal}{Phys. Rev. B}} \textbf{\bibinfo{volume}{91}},
  \bibinfo{pages}{054304} (\bibinfo{year}{2015}).

\bibitem{phono3py}
\bibinfo{author}{Togo, A.}, \bibinfo{author}{Chaput, L.} \&
  \bibinfo{author}{Tanaka, I.}
\newblock \bibinfo{title}{{Distributions of phonon lifetimes in Brillouin
  zones}}.
\newblock \emph{\bibinfo{journal}{Phys. Rev. B}} \textbf{\bibinfo{volume}{91}},
  \bibinfo{pages}{094306} (\bibinfo{year}{2015}).

\bibitem{paulatto2013anharmonic}
\bibinfo{author}{Paulatto, L.}, \bibinfo{author}{Mauri, F.} \&
  \bibinfo{author}{Lazzeri, M.}
\newblock \bibinfo{title}{{Anharmonic properties from a generalized third-order
  ab initio approach: Theory and applications to graphite and graphene}}.
\newblock \emph{\bibinfo{journal}{Phys. Rev. B}} \textbf{\bibinfo{volume}{87}},
  \bibinfo{pages}{214303} (\bibinfo{year}{2013}).

\bibitem{PhonTS}
\bibinfo{author}{Chernatynskiy, A.} \& \bibinfo{author}{Phillpot, S.~R.}
\newblock \bibinfo{title}{{Phonon Transport Simulator (PhonTS)}}.
\newblock \emph{\bibinfo{journal}{Comput. Phys. Commun.}}
  \textbf{\bibinfo{volume}{192}}, \bibinfo{pages}{196--204}
  (\bibinfo{year}{2015}).

\bibitem{alamode}
\bibinfo{author}{Tadano, T.}, \bibinfo{author}{Gohda, Y.} \&
  \bibinfo{author}{Tsuneyuki, S.}
\newblock \bibinfo{title}{{Anharmonic force constants extracted from
  first-principles molecular dynamics: applications to heat transfer
  simulations}}.
\newblock \emph{\bibinfo{journal}{J. Phys. Condens. Matter}}
  \textbf{\bibinfo{volume}{26}}, \bibinfo{pages}{225402}
  (\bibinfo{year}{2014}).

\bibitem{PhysRevB.96.014111}
\bibinfo{author}{Bianco, R.}, \bibinfo{author}{Errea, I.},
  \bibinfo{author}{Paulatto, L.}, \bibinfo{author}{Calandra, M.} \&
  \bibinfo{author}{Mauri, F.}
\newblock \bibinfo{title}{{Second-order structural phase transitions, free
  energy curvature, and temperature-dependent anharmonic phonons in the
  self-consistent harmonic approximation: Theory and stochastic
  implementation}}.
\newblock \emph{\bibinfo{journal}{Phys. Rev. B}} \textbf{\bibinfo{volume}{96}},
  \bibinfo{pages}{014111} (\bibinfo{year}{2017}).

\bibitem{PhysRevLett.120.105901}
\bibinfo{author}{Tadano, T.} \& \bibinfo{author}{Tsuneyuki, S.}
\newblock \bibinfo{title}{{Quartic Anharmonicity of Rattlers and Its Effect on
  Lattice Thermal Conductivity of Clathrates from First Principles}}.
\newblock \emph{\bibinfo{journal}{Phys. Rev. Lett.}}
  \textbf{\bibinfo{volume}{120}}, \bibinfo{pages}{105901}
  (\bibinfo{year}{2018}).

\bibitem{PhysRevB.94.020303}
\bibinfo{author}{van Roekeghem, A.}, \bibinfo{author}{Carrete, J.} \&
  \bibinfo{author}{Mingo, N.}
\newblock \bibinfo{title}{{Anomalous thermal conductivity and suppression of
  negative thermal expansion in ${\mathrm{ScF}}_{3}$}}.
\newblock \emph{\bibinfo{journal}{Phys. Rev. B}} \textbf{\bibinfo{volume}{94}},
  \bibinfo{pages}{020303} (\bibinfo{year}{2016}).

\bibitem{PhysRevB.98.024301}
\bibinfo{author}{Yang, F.~C.} \emph{et~al.}
\newblock \bibinfo{title}{{Temperature dependence of phonons in
  ${\mathrm{Pd}}_{3}\mathrm{Fe}$ through the Curie temperature}}.
\newblock \emph{\bibinfo{journal}{Phys. Rev. B}} \textbf{\bibinfo{volume}{98}},
  \bibinfo{pages}{024301} (\bibinfo{year}{2018}).

\bibitem{PhysRevLett.112.058501}
\bibinfo{author}{Zhang, D.-B.}, \bibinfo{author}{Sun, T.} \&
  \bibinfo{author}{Wentzcovitch, R.~M.}
\newblock \bibinfo{title}{{Phonon Quasiparticles and Anharmonic Free Energy in
  Complex Systems}}.
\newblock \emph{\bibinfo{journal}{Phys. Rev. Lett.}}
  \textbf{\bibinfo{volume}{112}}, \bibinfo{pages}{058501}
  (\bibinfo{year}{2014}).

\bibitem{PhysRevB.98.085205}
\bibinfo{author}{Ravichandran, N.~K.} \& \bibinfo{author}{Broido, D.}
\newblock \bibinfo{title}{{Unified first-principles theory of thermal
  properties of insulators}}.
\newblock \emph{\bibinfo{journal}{Phys. Rev. B}} \textbf{\bibinfo{volume}{98}},
  \bibinfo{pages}{085205} (\bibinfo{year}{2018}).

\bibitem{PhysRevB.91.214310}
\bibinfo{author}{Romero, A.~H.}, \bibinfo{author}{Gross, E. K.~U.},
  \bibinfo{author}{Verstraete, M.~J.} \& \bibinfo{author}{Hellman, O.}
\newblock \bibinfo{title}{{Thermal conductivity in PbTe from first
  principles}}.
\newblock \emph{\bibinfo{journal}{Phys. Rev. B}} \textbf{\bibinfo{volume}{91}},
  \bibinfo{pages}{214310} (\bibinfo{year}{2015}).

\bibitem{aseginolaza2018phonon}
\bibinfo{author}{Aseginolaza, U.} \emph{et~al.}
\newblock \bibinfo{title}{{Phonon Collapse and Second-Order Phase Transition in
  Thermoelectric SnSe}}.
\newblock \emph{\bibinfo{journal}{arXiv preprint arXiv:1807.07726}}
  (\bibinfo{year}{2018}).

\bibitem{PhysRevB.84.180301}
\bibinfo{author}{Hellman, O.}, \bibinfo{author}{Abrikosov, I.~A.} \&
  \bibinfo{author}{Simak, S.~I.}
\newblock \bibinfo{title}{{Lattice dynamics of anharmonic solids from first
  principles}}.
\newblock \emph{\bibinfo{journal}{Phys. Rev. B}} \textbf{\bibinfo{volume}{84}},
  \bibinfo{pages}{180301} (\bibinfo{year}{2011}).

\bibitem{PhysRevB.88.144301}
\bibinfo{author}{Hellman, O.} \& \bibinfo{author}{Abrikosov, I.~A.}
\newblock \bibinfo{title}{{Temperature-dependent effective third-order
  interatomic force constants from first principles}}.
\newblock \emph{\bibinfo{journal}{Phys. Rev. B}} \textbf{\bibinfo{volume}{88}},
  \bibinfo{pages}{144301} (\bibinfo{year}{2013}).

\bibitem{PhysRevB.87.104111}
\bibinfo{author}{Hellman, O.}, \bibinfo{author}{Steneteg, P.},
  \bibinfo{author}{Abrikosov, I.~A.} \& \bibinfo{author}{Simak, S.~I.}
\newblock \bibinfo{title}{{Temperature dependent effective potential method for
  accurate free energy calculations of solids}}.
\newblock \emph{\bibinfo{journal}{Phys. Rev. B}} \textbf{\bibinfo{volume}{87}},
  \bibinfo{pages}{104111} (\bibinfo{year}{2013}).

\bibitem{PhysRevLett.100.095901}
\bibinfo{author}{Souvatzis, P.}, \bibinfo{author}{Eriksson, O.},
  \bibinfo{author}{Katsnelson, M.~I.} \& \bibinfo{author}{Rudin, S.~P.}
\newblock \bibinfo{title}{{Entropy Driven Stabilization of Energetically
  Unstable Crystal Structures Explained from First Principles Theory}}.
\newblock \emph{\bibinfo{journal}{Phys. Rev. Lett.}}
  \textbf{\bibinfo{volume}{100}}, \bibinfo{pages}{095901}
  (\bibinfo{year}{2008}).

\end{thebibliography}

\end{document}